\documentclass[aps, prd, amsmath, floats, floatfix, twocolumn, nofootinbib, superscriptaddress]{revtex4-1}
\usepackage[pdftex]{graphicx}
\usepackage{color}
\usepackage{latexsym}
\usepackage[colorlinks]{hyperref}
\usepackage{amsthm,amsmath,amssymb}
\usepackage{mathrsfs}
\usepackage{makecell}

\newcommand{\be}{\begin{eqnarray}}
\newcommand{\ee}{\end{eqnarray}}

\begin{document}

\title{Testing the $\delta$-Kerr metric with black hole X-ray data}

\author{Jiahao~Tao}
\affiliation{Center for Field Theory and Particle Physics and Department of Physics, Fudan University, 200438 Shanghai, China}

\author{Shafqat~Riaz}
\affiliation{Center for Field Theory and Particle Physics and Department of Physics, Fudan University, 200438 Shanghai, China}
\affiliation{Theoretical Astrophysics, Eberhard-Karls Universit\"at T\"ubingen, D-72076 T\"ubingen, Germany}

\author{Biao~Zhou}
\affiliation{Center for Field Theory and Particle Physics and Department of Physics, Fudan University, 200438 Shanghai, China}

\author{Askar~B.~Abdikamalov}
\affiliation{School of Mathematics and Natural Sciences, New Uzbekistan University, Tashkent 100007, Uzbekistan}
\affiliation{Center for Field Theory and Particle Physics and Department of Physics, Fudan University, 200438 Shanghai, China}
\affiliation{Ulugh Beg Astronomical Institute, Tashkent 100052, Uzbekistan}

\author{Cosimo~Bambi}
\email[Corresponding author: ]{bambi@fudan.edu.cn}
\affiliation{Center for Field Theory and Particle Physics and Department of Physics, Fudan University, 200438 Shanghai, China}

\author{Daniele~Malafarina}
\affiliation{Department of Physics, Nazarbayev University, 010000 Astana, Kazakhstan}

\begin{abstract}
The spacetime around astrophysical black holes is thought to be described by the Kerr solution. However, even within general relativity, there is not yet a proof that the final product of the complete collapse of an uncharged body can only be a Kerr black hole. We can thus speculate on the possibility that the spacetime around astrophysical black holes may be described by other solutions of the Einstein Equations and we can test such a hypothesis with observations. In this work, we consider the $\delta$-Kerr metric, which is an exact solution of the field equations in vacuum and can be obtained from a non-linear superposition of the Kerr metric with a static axially symmetric solution, often referred to as the $\delta$-metric. The parameter $\delta=1+q$ quantifies the departure of the source from the Kerr metric and for $q=0$ we recover the Kerr solution. From the analysis of the reflection features in the X-ray spectrum of the Galactic black hole in EXO~1846--031, we find $-0.1 < q < 0.7$ (90\% CL), which is consistent with the hypothesis that the spacetime around the compact object in EXO~1846--031 is a Kerr black hole but does not entirely rule out the $\delta$-Kerr metric.  
\end{abstract}

\maketitle

%%%%%%%%%%%%%%%%%%%%%%%%%%%%%%%

\section{Introduction}

When a star exhausts all its nuclear fuel, the thermal pressure of the plasma cannot compensate the star's own weight and the body shrinks to find a new equilibrium configuration. If the collapsing part of the star exceeds the Oppenheimer-Volkoff limit, which is about 3~$M_\odot$ and corresponds to the maximum mass for a neutron star, there is no known mechanism to stop the collapse and we have the formation of a ``gravitationally collapsed object''~\cite{Ruffini:1969qy,Lattimer:2012nd}. It is generally assumed that the final outcome of complete collapse should be a black hole. However, the exact nature of such a gravitationally collapsed object is not yet completely understood and therefore theoretical and observational studies to address this question are active lines of research nowadays~\cite{Bambi:2015kza,Yagi:2016jml,Bambi:2017khi}.

In 4-dimensional general relativity, the only vacuum metric that is stationary, regular on and outside an event horizon, and asymptotically flat is the Kerr black hole solution. This is the celebrated result of a family of uniqueness theorems, which were pioneered in Refs.~\cite{Israel:1967wq,Carter:1971zc,Robinson:1975bv} and whose final version is still an ongoing research program~\cite{Chrusciel:2012jk}. In the '60s, Roger Penrose proposed the Cosmic Censorship Conjecture, according to which all singularities must be hidden behind an event horizon~\cite{Penrose:1969pc}. If this is true, within general relativity all gravitationally collapsed objects must be Kerr black holes. However, even within general relativity, the Cosmic Censorship Conjecture is still unproven and, at the same time, we know exact solutions of the Einstein Equations that violate the Cosmic Censorship Conjecture and in which the complete collapse of a body leads to a spacetime with naked singularities~\cite{Joshi:2011rlc}. 
A viewpoint that is widely accepted today regarding the appearance of singularities in solutions of Einstein's Equations is that they signal a regime where the theory fails and needs to be replaced by a new theory of gravity. In this sense, the study of singularities in general relativity may provide hints at the features that such a new theory must posses and how it may manifest in astrophysical phenomena~\cite{Malafarina:2017}.

From astrophysical observations, we know at least two classes of gravitationally collapsed objects: stellar-mass compact objects with a mass exceeding the Oppenheimer-Volkoff limit and supermassive objects in galactic nuclei. The latter are simply too massive, compact, and old to be clusters of non-luminous bodies like neutron stars~\cite{Maoz:1997yd}. For both object classes, we have even a body of observations suggesting that these objects do not have a normal surface but an event horizon~\cite{Narayan:2008bv,Broderick:2009ph}. The past few years have seen a tremendous progress in our capability of testing the nature of these compact objects and today we can use gravitational wave data from the LIGO-Virgo-KAGRA Collaboration~\cite{LIGOScientific:2016lio,Yunes:2016jcc,LIGOScientific:2019fpa,Cardenas-Avendano:2019zxd,Shashank:2021giy}, X-ray observations from a number of X-ray missions~\cite{Cao:2017kdq,Tripathi:2018lhx,Tripathi:2020dni,Tripathi:2020yts,Zhang:2021ymo,Tripathi:2021rqs,Bambi:2022dtw}, and the mm images of the supermassive objects in M87$^*$ and Sgr~A$^*$ from the Event Horizon Telescope Collaboration~\cite{Bambi:2019tjh,Wei:2020ght,Kumar:2020owy,EventHorizonTelescope:2022xqj,EventHorizonTelescope:2020qrl,Vagnozzi:2022moj}.

In the present paper, we want to explore the possibility that the spacetime around these gravitationally collapsed objects is described by the $\delta$-Kerr metric~\cite{deltakerr1,deltakerr2}, which is an exact solution of the Einstein Equations that can be obtained from a non-linear superposition of the $\delta$-metric (sometimes called Zipoy-Voorhees or $\gamma$-metric)\cite{ZV1,ZV2,ZV3,ZV4} and the Kerr metric. Such a solution, which can be understood as a stationary extension of the $\delta$-metric or a deformed extension of the Kerr metric, has three independent parameters: the mass parameter $M$, which is related to the mass of the compact object, the spin parameter $J$, which is related to the angular momentum of the source, and a deformation parameter, $q=1-\delta$, which quantifies the departure from the Kerr solution. For $q=0$ and $J\neq 0$, the $\delta$-Kerr metric reduces to the Kerr solution. For $J=0$ and $q\neq 0$, it reduces to the $\delta$-metric, while a non-vanishing value of $q$ and $J$ corresponds to a stationary, axisymmetric, and asymptotically flat vacuum solution of the Einstein Equations with a naked singularity.

The observational properties of the static $\delta$-metric have been widely studied in the past few years \cite{gamma1,gamma2,gamma3,gamma4,gamma5,gamma6,gamma7,gamma8}. However, since the $\delta$-metric is static it does not constitute a good candidate for the gravitational field outside an astrophysical source. On the other hand, the $\delta$-Kerr metric is stationary and continuously linked to the Kerr metric through the value of the deformation parameter and therefore it is an ideal candidate to test the validity of the Kerr hypothesis around astrophysical compact objects. The shadow of the $\delta$-Kerr spacetime was studied in Ref.~\cite{Li2022}, while its quasinormal modes were considered in Ref.~\cite{metric}.

The $\delta$-metric and the $\delta$-Kerr metric violate the Cosmic Censorship Conjecture, and for this reason they are normally not considered as viable solutions for the description of the spacetime around gravitationally collapsed objects.
Therefore in the absence of a proof of the Cosmic Censorship Conjecture and/or adopting the idea that quantum gravity effects may resolve spacetime singularities and make the Cosmic Censorship Conjecture unnecessary~\cite{Gimon:2007ur,Bambi:2008jg}, it is worth to check whether we can test and rule out the $\delta$-Kerr metric via astrophysical observations. 
To this aim, in this article we construct a reflection model for the $\delta$-Kerr metric and we analyze a \textsl{NuSTAR} spectrum of the X-ray binary EXO~1846--031 with strong reflection features. From the analysis of this observation, we can constrain the value of the deformation parameter $q$ of the source and thus test the $\delta$-Kerr spacetime.

The content of the paper is as follows. In Section~\ref{S2}, we briefly review the $\delta$-Kerr metric and, in Section~\ref{SXRS}, the analysis of the reflection features as a tool for testing the nature of gravitationally collapsed objects. In Section~\ref{S3}, we consider a \textsl{NuSTAR} observation of the X-ray binary EXO~1846--031 and we describe its data reduction. In Section~\ref{S4}, we present the spectral analysis of the \textsl{NuSTAR} observation and from the analysis of the reflection features we constrain the deformation parameter $q$ of the $\delta$-Kerr metric. Summary and conclusions are reported in Section~\ref{S5}. In the present manuscript, we adopt natural units with $G_{\rm N} = c = 1$ and the convention of a metric with signature $(-+++)$.

%%%%%%%%%%%%%%%%%%%%%%%%%%%%%%%

\section{$\delta$-Kerr metric \label{S2}}

Exact solutions of Einstein's field equations in vacuum describe the exterior of gravitating objects. There exist several classes of physically viable solutions besides the well known Schwarzschild and Kerr solutions. One of these is the so-called ``Weyl'' class that describes stationary, axially symmetric, vacuum solutions of the field equations \cite{weyl1,weyl2}.
In cylindrical coordinates $\{t,\rho,z,\phi\}$ the general line element of Weyl's class takes the form
\be 
ds^2&=&e^{-2\Lambda}\left[e^{2\Psi}(d\rho^2+dz^2)+\rho^2d\phi^2\right]+\\ \nonumber 
&&-e^{2\Lambda}(dt-\omega d\phi)^2,
\ee 
where $\Lambda=\Lambda(\rho,z)$, $\Psi=\Psi(\rho,z)$ and $\omega=\omega(\rho,z)$ are functions to be found from the field equations.
In the static case, i.e. for $\omega=0$, the field equations reduce to
\be \label{ee1}
0&=&\Lambda_{,\rho\rho}+\frac{\Lambda_{,\rho}}{\rho}+\Lambda_{,zz}, \\ \label{ee2}
\Psi_{,\rho}&=&\rho(\Psi_{,\rho}^2-\Psi_{,z}^2), \\ \label{ee3}
\Psi_{,z}&=&2\rho\Psi_{,\rho}\Psi_{,z},
\ee 
where we used the notation $\Lambda_{,x}=\partial \Lambda /\partial x$. Notice that \eqref{ee1} is nothing but the Laplace equation in flat space in cylindrical coordinates. Then once a solution of \eqref{ee1} is obtained, \eqref{ee2} and \eqref{ee3} are immediately integrated to obtain $\Psi$ with the integration constant fixed by requiring regularity of the metric at the center \cite{Quevedo1990}. 
Therefore there exists a one to one correspondence between solutions of Laplace equation and static solutions of the vacuum Einstein's equations in axial symmetry. One of the most well known solutions of the static equations is the $\delta$-metric \cite{ZV1,ZV2} which is given by
\be 
\Lambda&=& \frac{\delta}{2}\ln\left(\frac{R_++R_--2M}{R_++R_-+2M}\right) , \\
\Psi&=&\frac{\delta^2}{2}\ln\left(\frac{(R_++R_-)^2-4M^2}{4R_+R_-}\right),
\ee 
with
\be 
R_{\pm}=\rho^2+(z\pm M)^2.
\ee

The $\delta$-metric describes the field in the exterior of a static oblate or prolate object, with the parameter $\delta$ related to the departure from spherical symmetry. The Schwarzschild solution is obtained from the above for $\delta=1$ and it is the only solution of \eqref{ee1} describing a static black hole. In fact it is easy to show that for $\delta\neq 1$ the $\delta$-metric exhibits a curvature singularity at the location where the Schwarzschild metric has the event horizon, i.e. $r=2M$ in the spherical coordinates obtained from the coordinate change $\rho^2=(1-2M/r)r^2\sin^2\theta$ and $z=(r-M)\cos\theta$. Given the presence of a naked singularity at $r=2M$ one may question the physical interpretation of the spacetime in the whole range of allowed coordinates with $r>2M$. Since at present we do not know whether a new island of stability may exist for compact objects beyond the neutron star equation of state it is possible that the collapse of a non spherical object may settle to a non spherical final state without forming a black hole. Therefore it is worth to investigate the possibility that spacetimes such as the $\delta$-metric may describe exotic compact objects with boundary slightly larger than the Schwarzschild radius.

Of course there exist infinite static axially symmetric solutions of the field equations. This can be easily understood by thinking about the multipole expansion of the field of the exterior of a source \cite{geroch,hansen,hernandez}. The (spherical) Schwarzschild solution is unique and it corresponds to the choice of a non-vanishing monopole, i.e. $M$, and vanishing higher order multipoles at every order. Since there is an infinite number of terms in the multipole expansion, each choice of the set of multipole moments will yield a corresponding solution. The $\delta$-metric is of particular interest because all of the higher order multipole moments depend only on one continuous parameter $\delta$, which is directly related to the Schwarzschild solution and thus makes the interpretation of the geometry much easier \cite{ZV3,ZV4}. Of course astrophysical objects rotate. Therefore when considering model that should potentially describe astrophysical sources it is imperative to consider a non-vanishing angular momentum.

To obtain stationary solutions of the vacuum field equations one may apply some general techniques, such as the one proposed by Hoenselaers, Kinnersley and Xanthopoulos (HKX) \cite{HKX} or the Newman-Janis algorithm \cite{NJ}, which use known static solutions as `seeds' to obtain new corresponding stationary solutions.
For example, the Kerr metric can be obtained in this manner starting from the Schwarzschild solution and, as expected, it is the only black hole solution of the stationary vacuum equations. Similarly, one may take the $\delta$-metric as a seed to obtain a stationary generalization. However, one needs to be careful, since depending on the technique used one may obtain different, possibly non physically realistic, solutions for the stationary equations. For example, in \cite{halilsoy} the stationary metric obtained is a generalization of the $\delta$-metric to include a NUT-like parameter \cite{nut} and thus does not describe a rotating object. On the other hand, the $\delta$-Kerr metric considered in this paper describes the exterior of a rotating deformed object which reduces to the $\delta$-metric in the static case and to the Kerr metric in the case of no deformation. Similarly to the $\delta$-metric, the departure from the Kerr spacetime is entirely controlled by the value of one continuous parameter. This is a feature that makes the $\delta$-Kerr spacetime particularly interesting to study possible astrophysical applications. 
Of course there exist other stationary line elements that generalize Kerr and can be obtained from different static seed metrics. For example, one of the most widely known is the so-called Manko-Novikov spacetime \cite{manko}. However, the Manko-Novikov solution depends of the {\em ``full set of mass-multipole moments''}, which makes constraining its validity from observations much more complicated.

The $\delta$-Kerr metric was derived in Refs.~\cite{deltakerr1,deltakerr2} and can be thought of as a non-linear superposition of the $\delta$-metric and the Kerr metric. In Boyer-Lindquist-like coordinates $\left(t,r,\theta,\phi\right)$, the line element of the $\delta$-Kerr metric is~\cite{metric}
\begin{equation}
	\begin{split}
		\mathrm{d}s^2&=-F\mathrm{d}t^2+2F\omega\mathrm{d}t\mathrm{d}\phi+\frac{\mathrm{e}^{2\gamma}}{F}\frac{\mathbb{B}}{\mathbb{A}}\mathrm{d}r^2+\\
		&+r^2\frac{\mathrm{e}^{2\gamma}}{F}\mathbb{B}\mathrm{d}\theta^2
		+\left(\frac{r^2}{F}\mathbb{A}\sin^2\theta-F\omega^2\right)\mathrm{d}\phi^2\,,
		\label{metriceq}
	\end{split}
\end{equation}
where
\begin{equation}
	\begin{split}
		&\mathbb{A}=1-\frac{2M}{r}+\frac{a^2}{r^2}\,,\\
		&\mathbb{B}=\mathbb{A}+\frac{\sigma^2\sin^2\theta}{r^2}\,.
	\end{split}
\end{equation}
Here $M$ is the mass parameter, which is related to the gravitational mass of the compact object, $J$ is its spin parameter, related to the object's angular momentum, $a=J/M$ (while the dimensionless spin parameter is $a_*=a/M$), and $\sigma=\sqrt{M^2-a^2}>0$ is a constant length. $F$, $\omega$, and $\gamma$ are functions of the prolate spheroidal coordinates $x=\left(r-M\right)/\sigma$ and $y=\cos\theta$:
\begin{equation}
	\begin{split}
		&F=\frac{\mathcal{A}}{\mathcal{B}}\,,\\
		&\omega=2\left(a-\sigma\frac{\mathcal{C}}{\mathcal{A}}\right)\,,\\
		&\mathrm{e}^{2\gamma}=\frac{1}{4}\left(1+\frac{M}{\sigma}\right)^2\frac{\mathcal{A}}{\left(x^2-1\right)^{\delta}}\left(\frac{x^2-1}{x^2-y^2}\right)^{\delta^2}\,,
	\end{split}
\end{equation}
where
\begin{equation}
	\begin{split}
		\mathcal{A}&=a_+a_-+b_+b_-\,,\\
		\mathcal{B}&=a_+^2+b_+^2\,,\\
		\mathcal{C}&=\left(x+1\right)^q\left[x\left(1-y^2\right)\left(\lambda+\eta\right)a_++\right.\\
		&\left.+y\left(x^2-1\right)\left(1-\lambda\eta\right)b_+\right]\,;
	\end{split}
\end{equation}
\begin{equation}
	\begin{split}
		&a_\pm=\left(x\pm1\right)^q\left[x\left(1-\lambda\eta\right)\pm\left(1+\lambda\eta\right)\right]\,,\\
		&b_\pm=\left(x\pm1\right)^q\left[y\left(\lambda+\eta\right)\mp\left(\lambda-\eta\right)\right]\,;
	\end{split}
\end{equation}
\begin{equation}
	\begin{split}
		&\lambda=\alpha\left(x^2-1\right)^{-q}\left(x+y\right)^{2q}\,,\\
		&\eta=\alpha\left(x^2-1\right)^{-q}\left(x-y\right)^{2q}\,;
	\end{split}
\end{equation}
\begin{equation}
	\begin{split}
		&q=\delta-1\,,\\
		&\alpha=\frac{M-\sigma}{a}=\frac{a}{M+\sigma}\,.
	\end{split}
\end{equation}

Contrary to the Schwarzschild solution, for $q\neq 0$ the $\delta$-Kerr spacetime has a non-vanishing mass quadrupole moment even when $a=0$ \cite{Li2022}.
For $q=0$, we have $\lambda=\eta=\alpha$, $a_{\pm}=2\alpha\left(r-M\pm M\right)/a$, and $b_{\pm}=2\alpha \cos\theta$. Therefore, we can get
\begin{equation}
	\begin{split}
		&\mathcal{A}=\frac{4\alpha^2}{a^2}\left(\Sigma-2Mr\right)\,,\\
		&\mathcal{B}=\frac{4\alpha^2}{a^2}\Sigma\,,\\
		&\mathcal{C}=\frac{4\alpha^2}{a\sigma}\left[\Sigma-Mr\left(1+\cos^2\theta\right)\right]\,,
	\end{split}
\end{equation}
and
\begin{equation}
	\begin{split}
		&F=1-\frac{2Mr}{\Sigma}\,,\\
		&\omega=-\frac{2Mar\sin^2\theta}{\Sigma-2Mr}\,,\\
		&\mathrm{e}^{2\gamma}=\frac{\Sigma-2Mr}{\Sigma-2Mr+M^2\sin^2\theta}\,,
	\end{split}
\end{equation}
where $\Sigma=r^2+a^2\cos^2\theta$. If we plug these expressions in the line element in Eq.~(\ref{metriceq}), we recover the familiar Kerr solution in Boyer-Lindquist coordinates. Finally note that for $a=0$ and $q=0$ we retrieve the Schwarzschild metric.

In order to relate the parameters $M$, $a$ and $q$ to measurable quantities, it is useful to consider the multipole expansion of the mass and angular momentum of the source~\cite{Quevedo1990}. Looking only at the first two non-vanishing multipoles, we have that the mass monopole and mass quadrupole are
\be
 M_0&=& M +q \sqrt{M^2-a^2} \ , \\ 
 M_2&=& \frac{\sigma^3 q}{3}(7-q^2)+M\sigma^2(1-q^2)-M^2(M+3\sigma q) \ , \nonumber\\
\ee
while for the angular momentum we have
\be 
 J_1&=& aM+2aq\sqrt{M^2-a^2} \ ,  \\ 
 J_3&=& -a^3M-aM\sigma q(3\sigma q+4M)-\frac{2}{3} a \sigma^3 q (q^2-4) \ . \nonumber\\
\ee
Notice that they reduce to the values for the Zipoy-Voorhees metric in the non rotating case $a=0$, and they reduce to the values for Kerr when $q=0$.

From the above discussion it should be clear that the $\delta$-Kerr metric represents an ideal candidate to test possible deviations from the Kerr metric of the geometry in the surroundings of extreme compact objects. In particular it may be possible to devise measurements to estimate the value of $\delta$ for real astrophysical black hole candidates and thus constrain, confirm or exclude the physical validity of such solutions.

%%%%%%%%%%%%%%%%%%%%%%%%%%%%%%%

\section{X-ray reflection spectroscopy \label{SXRS}}

\begin{figure}[t]
\centering
\includegraphics[width=0.48\textwidth]{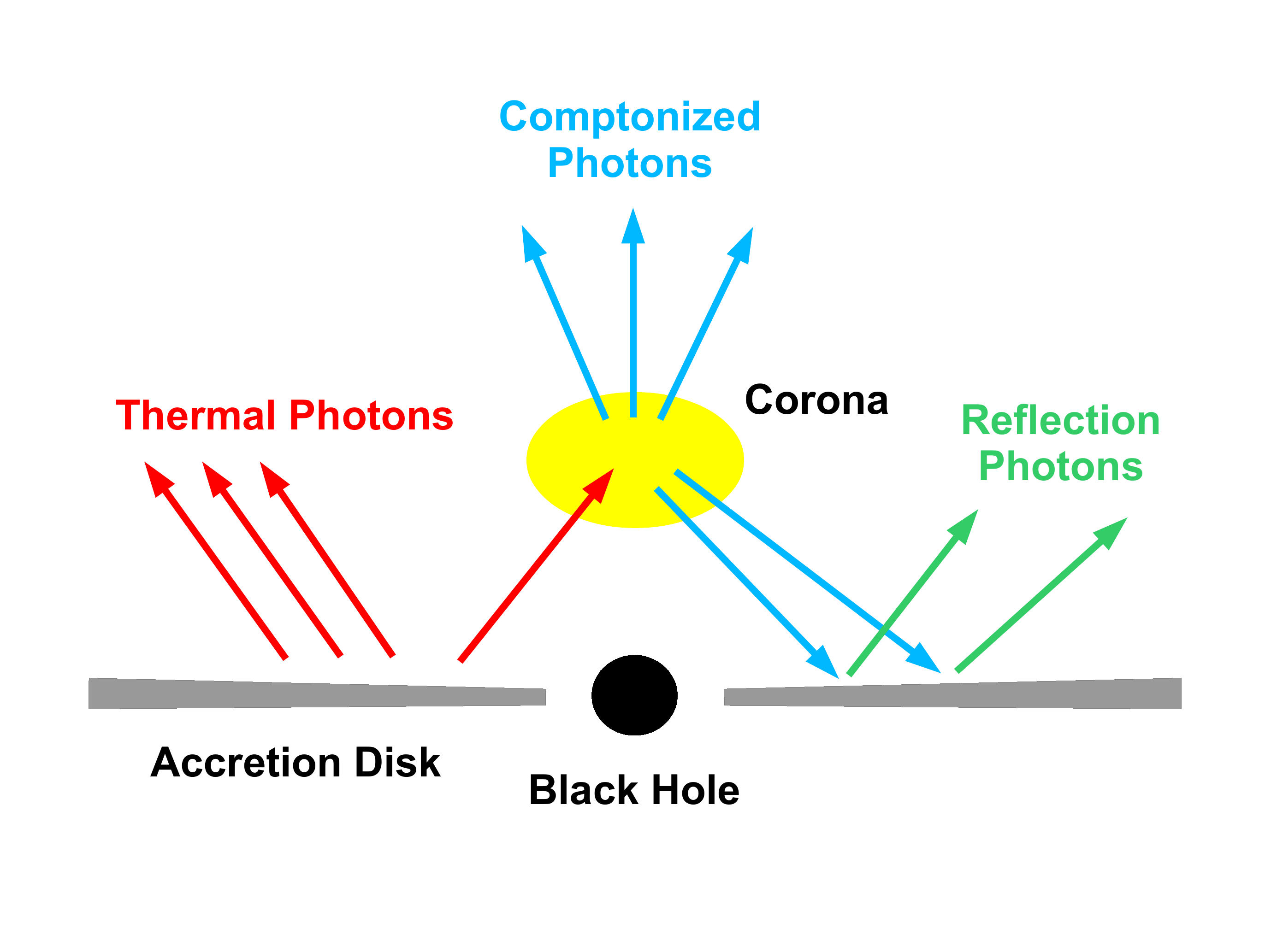}
\vspace{-1.0cm}
\caption{Disk-corona system. Figure from Ref.~\cite{Bambi:2021chr} under the terms of the Creative Commons Attribution 4.0 International License.}
\label{f-corona}
\end{figure}

Relativistically blurred reflection features are common in the X-ray spectra of accreting black holes~\cite{Tanaka:1995en,Nandra:2007rp,Miller:2009cw}. These features are produced by illumination of a ``cold'' disk by a ``hot'' corona~\cite{Bambi:2020jpe}. The astrophysical system is shown in Fig.~\ref{f-corona}. The accretion disk around the black hole is optically thin and geometrically thick. The gas in the disk is in local thermal equilibrium\footnote{We note that this is a common assumption in current theoretical models. In reality, non-equilibrium conditions may exist.} and every point on the surface of the disk emits a blackbody-like spectrum. The whole disk has a multi-temperature blackbody-like spectrum because the temperature of the gas increases approaching the central object. The thermal spectrum of the accretion disk is normally peaked in the soft X-ray band (0.1-10~keV) in the case of stellar-mass black holes in X-ray binary systems and in the UV band (1-100~eV) in the case of supermassive black holes in active galactic nuclei. The corona is some hotter plasma ($\sim 100$~keV) near the black hole. Thermal photons from the accretion disk can inverse Compton scatter off free electrons in the corona. The Comptonized photons can illuminate the disk: Compton scattering and absorption followed by fluorescent emission generate the reflection spectrum.

In the rest-frame of the gas in the disk, the reflection spectrum is characterized by narrow fluorescent emission lines below 10~keV and a Compton hump peaking at 20-30~keV~\cite{Ross:2005dm,Garcia:2010iz}. The reflection spectrum of the whole disk detected by a distant observer is blurred because it is the result of photons coming from all points of the accretion disk and every point of the disk is characterized by its own redshift factor, resulting from the combination of gravitational redshift and Doppler boosting~\cite{Bambi:2017khi}. X-ray reflection spectroscopy refers to the analysis of the reflection features in the X-ray spectra of accreting black holes. In the presence of high-quality data and the correct astrophysical model, X-ray reflection spectroscopy can be a powerful technique to probe the strong gravity regions around black hole candidates~\cite{Bambi:2020jpe}.

The idea of using the analysis of reflection features to test the nature of gravitationally collapsed objects and the Kerr black hole hypothesis was discussed for the first time in Ref.~\cite{Lu:2002vm} and further explored by other authors in Refs.~\cite{Schee:2008fc,Johannsen:2012ng,Bambi:2012at,Bambi:2013jda,Bambi:2013hza}. In those early works, it was only studied the shape of the iron K$\alpha$ line, which is often one of the most prominent features in the reflection spectrum and certainly its most informative part about the spacetime metric in the strong gravity region around the compact object. However, none of those models was suitable to analyze real data. A breakthrough in this field was the development of the reflection model {\tt relxill\_nk}~\cite{Bambi:2016sac,Abdikamalov:2019yrr,Abdikamalov:2020oci}, which is an extension of the popular {\tt relxill} package~\cite{Dauser:2013xv,Garcia:2013lxa} for non-Kerr spacetimes. {\tt relxill\_nk} has been extensively used in the past few years to test the Kerr black hole hypothesis and specific modified theories of gravity in which rotating compact objects are not described by the Kerr solution (e.g.~\cite{Zhou:2018bxk,Zhu:2020cfn,Zhou:2020eth,Riaz:2022rlx}). The state-of-the-art in the field is reviewed in Ref.~\cite{Bambi:2022dtw}.

In general, the observed flux of an accretion disk around a compact object can be calculated as
\be\label{eq-flux}
F (E_{\rm o}) &=& \frac{1}{D^2} \int dX \, dY \, I_{\rm o} (X,Y)= \nonumber\\
&=& \frac{1}{D^2} \int dX \, dY \, g^3 I_{\rm e} (E_{\rm o}, r_{\rm e} , \vartheta_{\rm e}) \, , 
\ee
where $I_{\rm o}$ and $I_{\rm e}$ are the specific intensity of the radiation as measured, respectively, by the distant observer and in the rest-frame of the gas in the disk. $X$ and $Y$ are the Cartesian coordinates of the image of the disk in the plane of the distant observer and $D$ is the distance of the observer from the source. $I_{\rm o} = g^3 I_{\rm e}$ follows from Liouville's theorem, $g = E_{\rm o}/E_{\rm e}$ is the redshift factor, and $E_{\rm o}$ and $E_{\rm e}$ are the photon energies as measured, respectively, by the distant observer and in the rest-frame of the gas. Here $r_{\rm e}$ is the emission radius on the disk and $\vartheta_{\rm e}$ is the emission angle, which may differ from the inclination angle of the disk with respect to the line of sight of the distant observer, $i$, because of light bending. The natural way to calculate the observed flux $F (E_{\rm o})$ is to consider a grid on the plane of the distant observer and follow the trajectories of photons backwards in time from every point of the grid to the disk~\cite{Reynolds:1997ek,Wilkins:2016qqz}. In this way, we connect every point of the image of the disk on the plane of the distant observer with its actual emission point on the disk, we can calculate the redshift factor $g$, and, if we know the local spectrum, we can calculate the integral.

\begin{figure*}[t]
	\centering
	\includegraphics[width=0.95\textwidth,trim=0.0cm 0.0cm 1.5cm 0.0cm,clip]{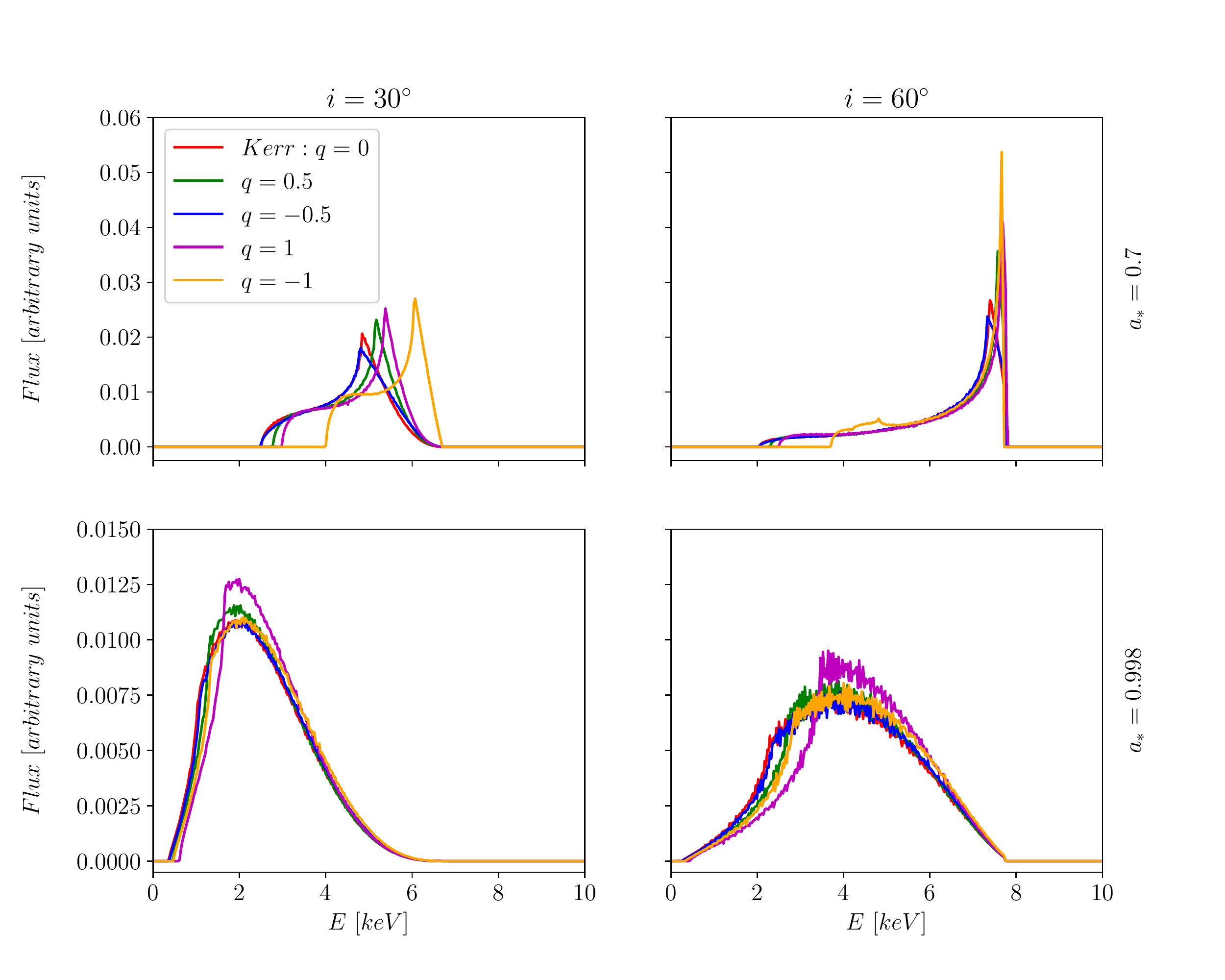}
	\vspace{-0.6cm}
	\caption{Iron line profiles in $\delta$-Kerr spacetimes. The inclination angle of the disk with respect to the line of sight of the distant observer is $i = 30^\circ$ (left panels) and $60^\circ$ (right panels). The dimensionless spin parameter is $a_* = 0.7$ (top panels) and 0.998 (bottom panels). These profiles are calculated assuming that the emissivity profile is described by a power law with emissivity index $p=8$, the inner edge of the disk is at the ISCO radius, and the outer edge is at 400~$r_{\rm g}$, where $r_{\rm g}$ is the gravitational radius. \label{ilp}}	 
\end{figure*}

In practice, this approach is not doable because the ray-tracing calculations are too time consuming to be done during the data analysis process. The current strategy in most reflection models, including also {\tt relxill\_nk}, is to introduce the ``transfer function'' and rewrite Eq.~(\ref{eq-flux}) as (see, e.g., Refs.~\cite{Bambi:2017khi,Bambi:2016sac})
\be
F (E_{\rm o}) &=& \frac{1}{D^2} \int_{r_{\rm in}}^{r_{\rm out}} dr_{\rm e} \int_0^1 dg^* \frac{\pi r_{\rm e} g^2}{\sqrt{g^* (1 - g^*)}} \nonumber\\ && \qquad \qquad \times f (g^* , r_{\rm e} , i) \, I_{\rm e} (E_{\rm o}, r_{\rm e} , \vartheta_{\rm e}) \, ,
\ee
where $r_{\rm in}$ and $r_{\rm out}$ are, respectively, the inner and the outer edges of the disk, $f$ is the transfer function~\cite{Cunningham:1975zz}
\be
f (g^* , r_{\rm e} , i) = \frac{g \sqrt{g^* (1 - g^*)}}{\pi r_{\rm e}} J (X,Y; g^* , r_{\rm e}) \, ,
\ee
$g^*$ is the relative redshift at the emission radius $r_{\rm e}$ for an observer with viewing angle $i$
\be
g^* = \frac{g - g_{\rm min}}{g_{\rm max} - g_{\rm min}} \, ,
\ee
and $g_{\rm min} = g_{\rm min} (r_{\rm e},i)$ and $g_{\rm max} = g_{\rm max} (r_{\rm e},i)$ are, respectively, the minimum and the maximum values of the redshift factor $g$ for photons emitted at the emission radius $r_{\rm e}$ and detected by an observer with viewing angle $i$. Finally $J (X,Y; g^* , r_{\rm e})$ is the Jacobian between the Cartesian coordinates of the image of the disk in the plane of the distant observer and the two variables $r_{\rm e}$ and $g^*$ used to map the emission points on the accretion disk.

The transfer function and the non-relativistic reflection spectrum can be calculated before the data analysis process on a computer cluster and tabulated in FITS files for a grid of their input parameters. During the data analysis process, the model calls the FITS files and can quickly calculate the integral to obtain the observed spectrum. If we want to construct a model for a different spacetime metric, we just need to replace the old FITS file of the transfer function with a new one, which is calculated for the new metric of interest. This is what we have done to implement the $\delta$-Kerr metric in {\tt relxill\_nk}: we have considered a grid of spin parameters $a_*$, viewing angles $i$, and deformation parameters $q$ and for every point of the grid we have calculated the transfer function with a ray-tracing code for 100 values of $r_{\rm e}$ and 40 values of $g^*$.

If, instead of the full non-relativistic reflection spectrum, we consider only a narrow iron line at 6.4~keV in $I_{\rm e}$, the calculation of $F (E_{\rm o})$ produces relativistically broadened iron lines. While any precise measurement from the analysis of the reflection features requires to consider the full reflection spectrum and not only an iron line, a single iron line can show better the impact of the parameter $q$ on the shape of the spectrum. Fig.~\ref{ilp} shows some relativistically broadened iron lines for two values of the inclination angle ($i = 30^\circ$ and $60^\circ$) and two values of the spin parameter ($a_* = 0.7$ and $0.998$). For every line, the emissivity profile is supposed to be a power law with emissivity index $p=8$; in reality, a very steep emissivity profile can only be expected in the inner part of the accretion disk~\cite{Wilkins:2012zm}, but here we want to show the impact of the deformation parameter $q$ on an iron line profile and we consider a simple case that maximizes the effect. The inner edge of the disk is set at the radius of the innermost stable circular orbit (ISCO) and the outer edge of the disk is set at 400 gravitational radii. In every panel, we show the iron line for $q=0$ (Kerr spacetime), $\pm 0.5$, and $\pm 1$. Notice that the value $q=-1$, corresponding to $\delta=0$ is the limiting case of an extremely flattened source, which in the static case corresponds to the Curzon solution \cite{Curzon}.

%%%%%%%%%%%%%%%%%%%%%%%%%%%%%%%

\section{Observation and data reduction \label{S3}}

\subsection{Selection of the source}

With the $\delta$-Kerr metric implemented in {\tt relxill\_nk}, we can select reflection dominated spectra of accreting black holes and fit these spectra to measure the deformation parameter $q$ of the selected sources. Since our goal is to test fundamental physics, not to study the astrophysical environment of specific sources, we need to select sources and observations that can permit us to get robust measurements of the spacetime geometry. This issue is already discussed in the literature; see, for example, Ref.~\cite{Bambi:2022dtw} for tests of the Kerr hypothesis and Refs.~\cite{Dauser:2013xv,Kammoun:2018ltv} for black hole spin measurements, but the conclusions are the same. It turns out that it is {\it extremely} important to select suitable sources and observations. We clearly need to select spectra with strong reflection features, but it is also crucial that most of the reflection component is generated very close to the black hole in order to maximize the relativistic effects in the spectrum. This requires to select sources in which the inner edge of the disk is very close to the black hole (which, in turn, requires to select very-fast rotating sources) and that the emissivity profile in the inner part of the accretion disk is very steep. It is also necessary to select sources with geometrically thin accretion disks~\cite{Shashank:2021giy,Riaz:2019bkv,Riaz:2019kat} (which, in turn, requires sources with an accretion luminosity lower than about 30\% of their Eddington limit), bright (to have a good statistics), with constant flux and hardness (otherwise, the morphology of the accretion flow and/or the corona may change during the observation), and the data should not be affected by pile-up. It is also very important to have data covering the whole X-ray spectrum (and not only the iron line region).

From past studies, we know the sources and observations most suitable to test the spacetime metric using X-ray reflection spectroscopy. From the sole analysis of the reflection features, the spectrum of EXO~1846--031 observed by \textsl{NuSTAR} on August 3, 2019 is certainly one of the best options~\cite{Bambi:2022dtw}, nicely meeting the selection criteria listed above. The source was in a hard intermediate state, with very strong reflection features, inner edge of the disk very close to the central object, steep emissivity profile, and quite a high count rate. The analysis of the reflection features in this spectrum can provide very stringent constraints on the spacetime metric around EXO~1846--031 and systematic effects seem to be under control, so any measurement is also accurate.

\subsection{EXO~1846--031}

EXO~1846--031 is a low mass X-ray binary~\cite{exo1}. It was discovered by the European X-ray Observatory Satellite (\textsl{EXOSAT}) on April 3, 1985~\cite{exo2}. A second outburst was detected by \textsl{CGRO}/BATSE in 1994~\cite{exo3}. After being in quiescence for about 25~years, the source had a new outburst in 2019, which was first detected by \textsl{MAXI} on July 23~\cite{exo4}. This third outburst was then observed by other instruments; e.g., \textsl{Swift}/XRT~\cite{exo5}, VLA~\cite{exo6}, and MeerKAT~\cite{exo7}. The Nuclear Spectroscopic Telescope Array mission (\textsl{NuSTAR})~\cite{NuSTAR} observed EXO~1846--031 on August 3, 2019 (observation ID 90501334002) with a 22.2 ks exposure time. In what follows, we will consider this \textsl{NuSTAR} observation, which was first analyzed in Ref.~\cite{exoN}.

\subsection{Data reduction}

For the data reduction, we follow Ref.~\cite{exoN}. \textsl{NuSTAR} has two detectors, which are called Focal Plane Modules (FPM) A and B. We download the raw data from the HEASARC website and use the HEASOFT module {\tt nupipeline} to convert the raw data into cleaned event files with NuSTARDAS and the CALDB 20220301 calibration database, so that we can get the source and background information. For the source, we select a 180~arcseconds radius circular region at the center of the source for both FPMA and FPMB. For the background, we take a region of the same size of the source as far as possible from the source but on the same detector, so that the influence of the source's photons can be ignored. Afterwards, we use the HEASOFT module {\tt nuproducts} to generate the source and background spectra, the response matrix file, and the ancillary file. Last, we use {\tt grppha} to group the spectra to have at least 30~counts per bin. Since the new CALDB corrects the calibration in the 3-7~keV energy range, we do not need the table {\tt nuMLIv1.mod} used in Ref.~\cite{exoN}.

%%%%%%%%%%%%%%%%%%%%%%%%%%%%%%%

\section{Spectral analysis \label{S4}}

\begin{figure}[htbp]
	\centering
	\includegraphics[width=0.48\textwidth,trim=1.0cm 0.0cm 1.5cm 0.0cm,clip]{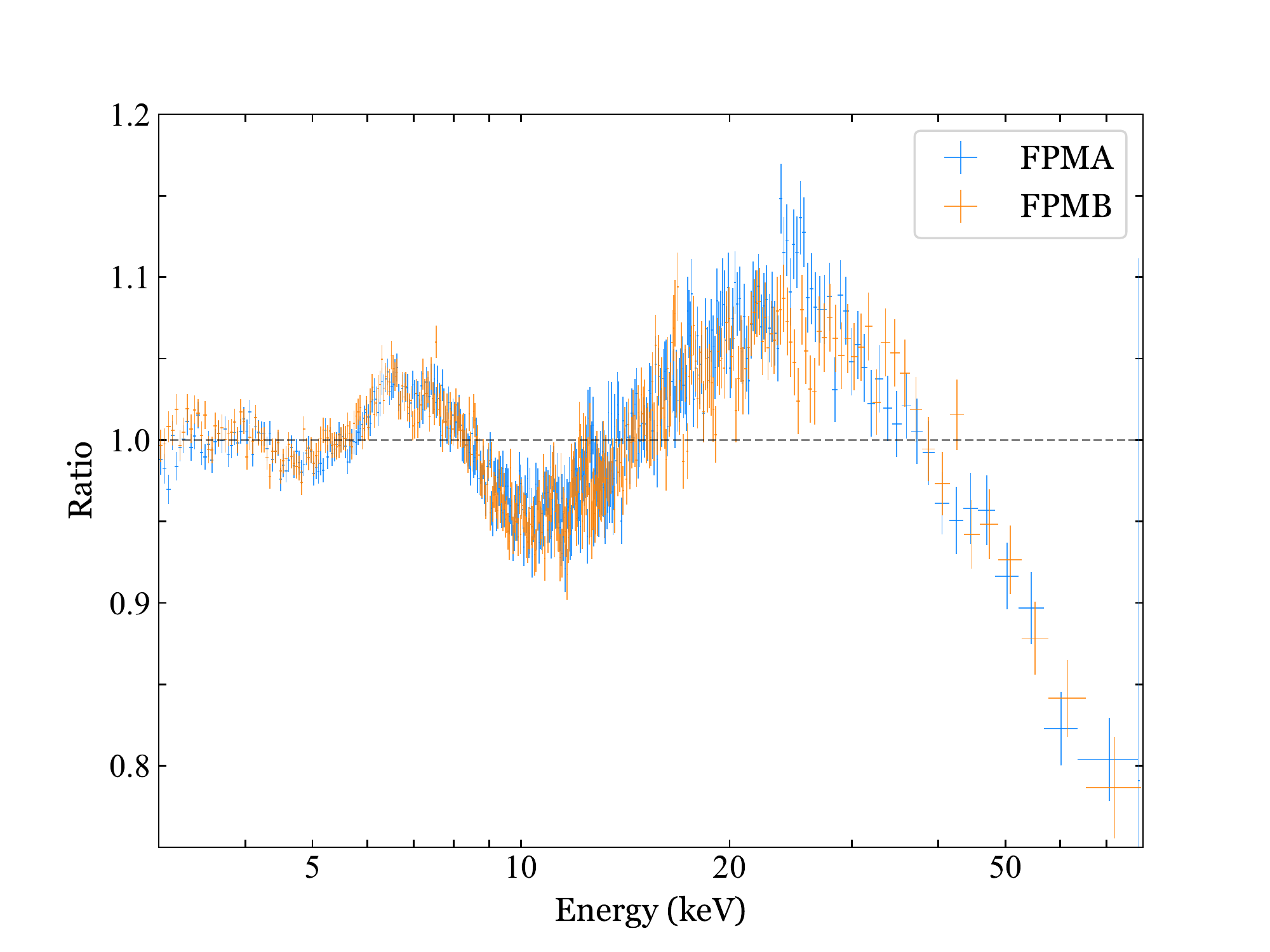}
	\vspace{-0.6cm}
	\caption{Data to best-fit model ratio for an absorbed power law. We clearly see a broadened iron line peaking around 7~keV and a Compton hump peaking at 20-30~keV. \label{ratio}}	 
\end{figure}

For the spectral analysis, we use XSPEC v12.12.1~\cite{xspec}. First, we fit the data with an absorbed power law to see the reflection features in the spectrum. In XSPEC language, the model reads 
\begin{equation*}
	{\tt const\times tbabs\times (diskbb+cutoffpl)} \, .
\end{equation*}
{\tt const} is used to have a cross-calibration constant between the detectors FPMA and FPMB: the constant is frozen to 1 for FPMA and is free for FPMB. {\tt tbabs} describes the Galactic absorption~\cite{tbabs}: the hydrogen column density, $N_\mathrm{H}$, is the only parameter of the model and is left free in the fit. {\tt diskbb} describes the thermal spectrum of the accretion disk~\cite{diskbb}: the temperature at the inner edge of the disk, $T_\mathrm{in}$, and the normalization of the component are left free in the fit. {\tt cutoffpl} describes the continuum from the corona: the photon index, $\Gamma$, the high energy cutoff, $E_\mathrm{cut}$, and the normalization of this component are left free in the fit. The ratio between the data and the best-fit model is shown in Fig.~\ref{ratio} and we clearly see unresolved strong reflection features: a broadened iron K$\alpha$ line peaking around 7~keV and a Compton hump peaking at 20-30~keV. Such a strong blurred reflection features suggest that this \textsl{NuSTAR} spectrum is suitable to test the nature of the gravitationally collapsed object in EXO~1846--031 with {\tt relxill\_nk}.

To fit the reflection features, we add {\tt relxill\_nk} to the total model. We employ the flavor {\tt relxillion\_nk}, which describes the relativistically blurred reflection spectrum of an accretion disk with a non-trivial ionization gradient~\cite{Abdikamalov:2021rty,Abdikamalov:2021ues}. In XSPEC language, the total model now reads  
\begin{equation*}
	{\tt const\times tbabs\times (diskbb+relxillion\_nk)} \, .
\end{equation*}
{\tt relxill\_nk} has several parameters. The spacetime metric is described by the spin $a_*$ and the deformation parameter $q$ of the $\delta$-Kerr metric and both parameters are left free in the fit. The inner edge of the accretion disk is set at the ISCO and therefore it is not a free parameter but directly depends on the values $a_*$ and $q$. The outer edge of the disk is fixed to 900~$r_\mathrm{g}$, where $r_\mathrm{g}$ is the gravitational radius and 900~$r_\mathrm{g}$ is the maximum value allowed by the model. The emissivity profile of the accretion disk can potentially be described by a twice broken power law and there are thus five parameters: the emissivity indices of the inner, central, and outer regions ($p_1$, $p_2$, and $p_3$, respectively) and the breaking radii between the inner and the central parts, $R_\mathrm{br1}$, and between the central and outer parts, $R_\mathrm{br2}$\footnote{This means that the emission of the disk scales as $r^{-p_1}$ in the inner part ($r < R_\mathrm{br1}$), as $r^{-p_2}$ in the central part ($R_\mathrm{br1} < r < R_\mathrm{br2}$), and as $r^{-p_3}$ in the outer part ($r > R_\mathrm{br2}$).}. To model the emissivity profile with a broken power law (instead of a twice broken power law), we simply set $p_2 = p_3$ and $R_\mathrm{br1} = R_\mathrm{br2}$ (i.e., the central region collapses and we have only the inner and outer regions). The viewing angle, $i$, the iron abundance, $A_{\rm Fe}$, the ionization at the inner edge of the disk, $\xi_{\rm in}$, and the ionization index, $\alpha_\xi$, are all free parameters in the fit. The model includes the continuum from the corona and the reflection fraction, $R_{\rm f}$, regulates the relative strength between the reflection component and the continuum. The photon index, $\Gamma$, and the high-energy cutoff, $E_\mathrm{cut}$, of the continuum illuminating the disk are free in the fit.

\begin{table*}
\renewcommand\arraystretch{1.5}
	\centering
	\vspace{0.5cm}
	\begin{tabular}{ccc}
		\hline\hline
		\hspace{0.5cm} Model \hspace{0.5cm} & XSPEC Model & \hspace{0.5cm} Emissivity Profile \hspace{0.5cm} \\
		\hline\hline
		1 & {\tt tbabs$\times$(diskbb+relxillion\_nk)} & \makecell{$p_1$, $p_2$, $R_{\rm br 1}$}\\
		\hline
		2 & {\tt tbabs$\times$(diskbb+relxillion\_nk)} & \makecell{$p_1$, $p_2$, $p_3=3$, $R_{\rm br 1}$, $R_{\rm br 2}$}\\
		\hline
		3 & {\tt tbabs$\times$(diskbb+relxillion\_nk)} & \makecell{$p_1$, $p_2 = 3$, $R_{\rm br 1}$}\\
		\hline
		4 & {\tt tbabs$\times$(diskbb+relxill\_nk+xillver)} & \makecell{$p_1$, $p_2$, $R_{\rm br 1}$}\\
		\hline\hline
	\end{tabular}
	%\vspace{0.2cm}
	\caption{Summary of the models used in the spectral analysis of this work.\label{models}}
\end{table*}

From previous analyses~\cite{Abdikamalov:2021rty,Abdikamalov:2021ues}, we know that this spectrum requires a non-vanishing ionization gradient and for this reason we use the flavor {\tt relxillion\_nk}. If we fit the data with a model with a disk with constant ionization, we need to add a Gaussian to the total model~\cite{exoN}. We fit the data with four models (Models~1-4), which are listed in Tab.~\ref{models}.

\begin{figure*}[t]
	\begin{center}
		\includegraphics[width=0.48\textwidth,trim=0.8cm 0.0cm 1.0cm 0.0cm,clip]{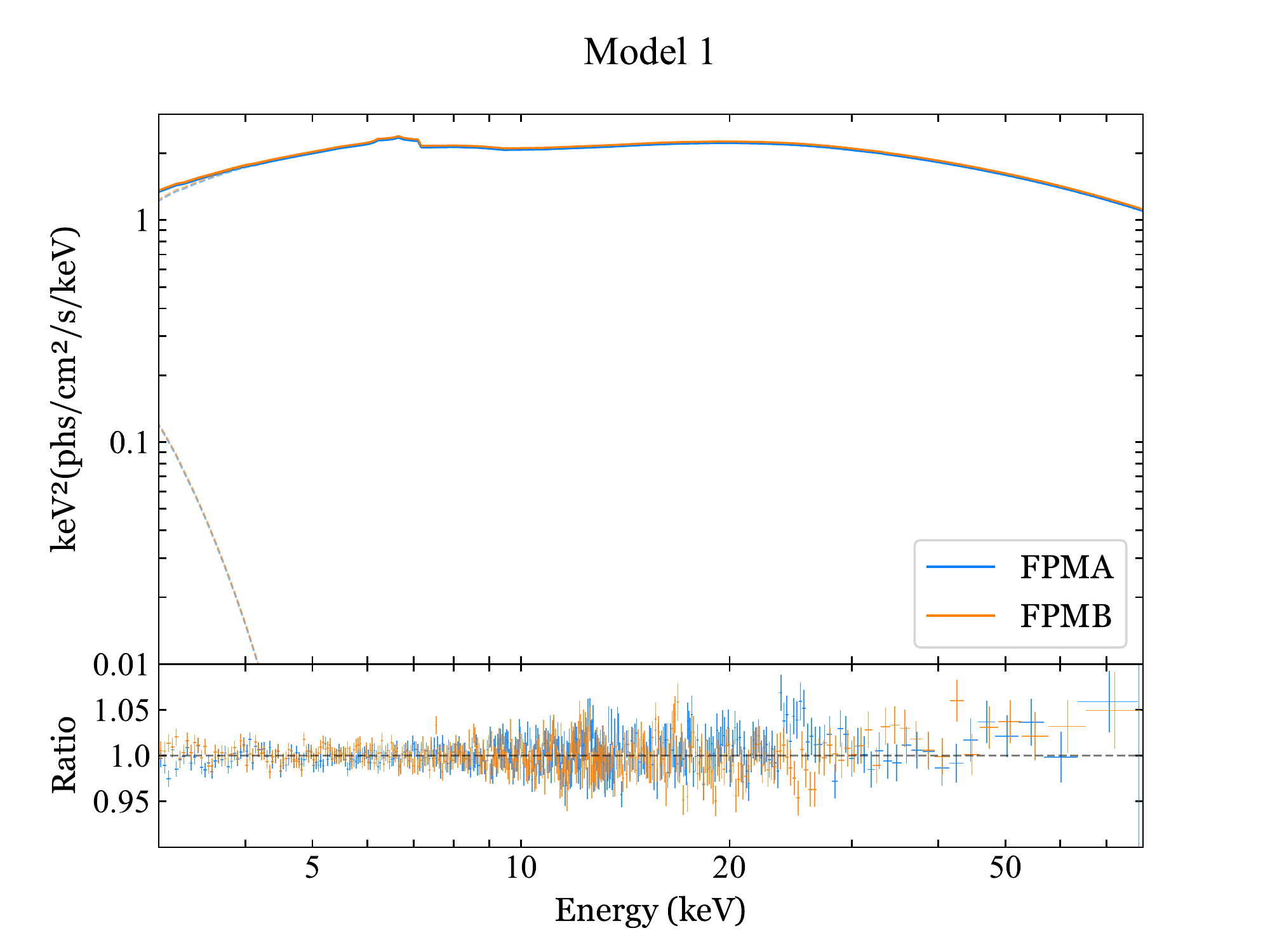}
		\includegraphics[width=0.48\textwidth,trim=0.8cm 0.0cm 1.0cm 0.0cm,clip]{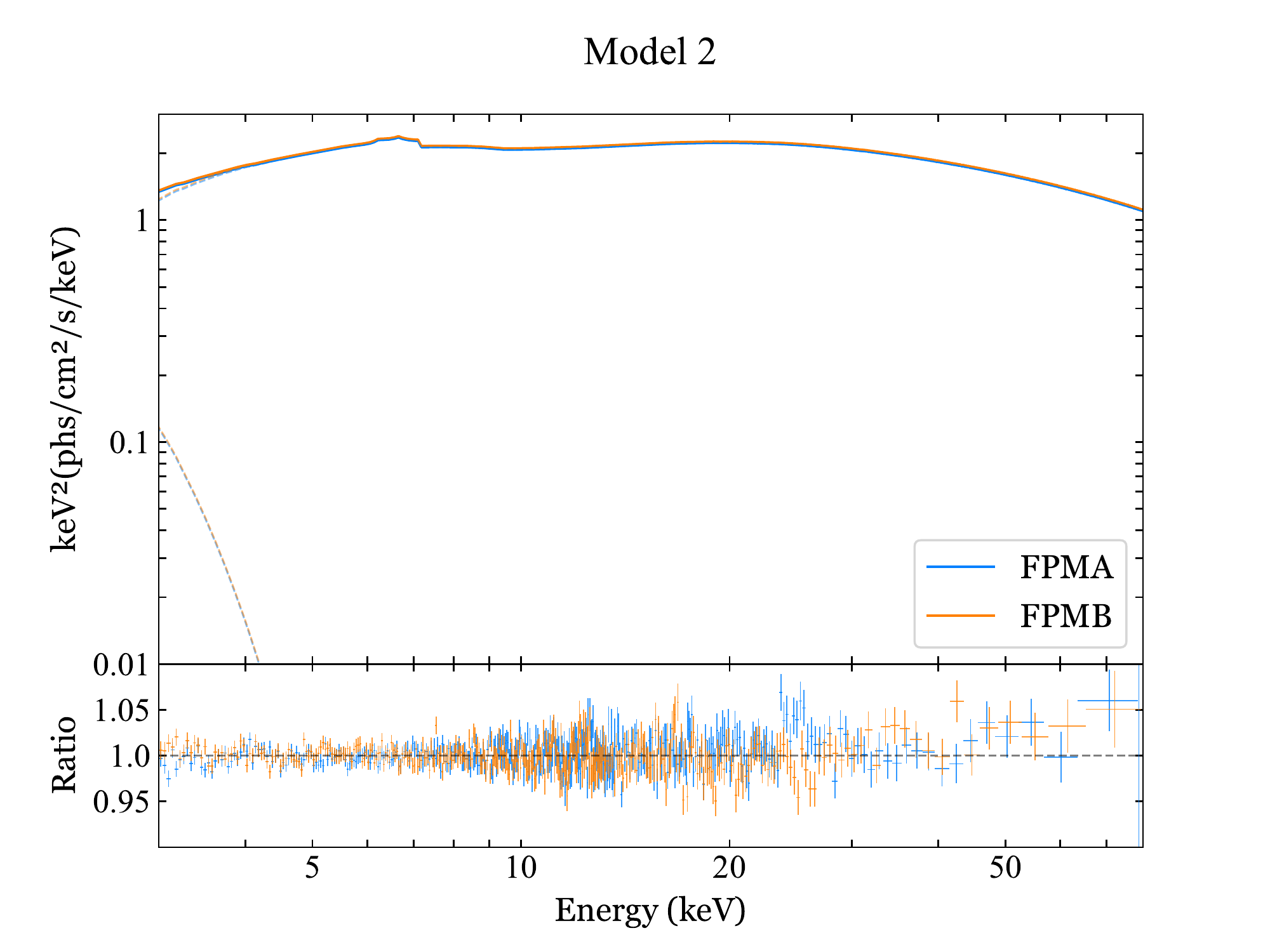} \\ \vspace{0.2cm}
		\includegraphics[width=0.48\textwidth,trim=0.8cm 0.0cm 1.0cm 0.0cm,clip]{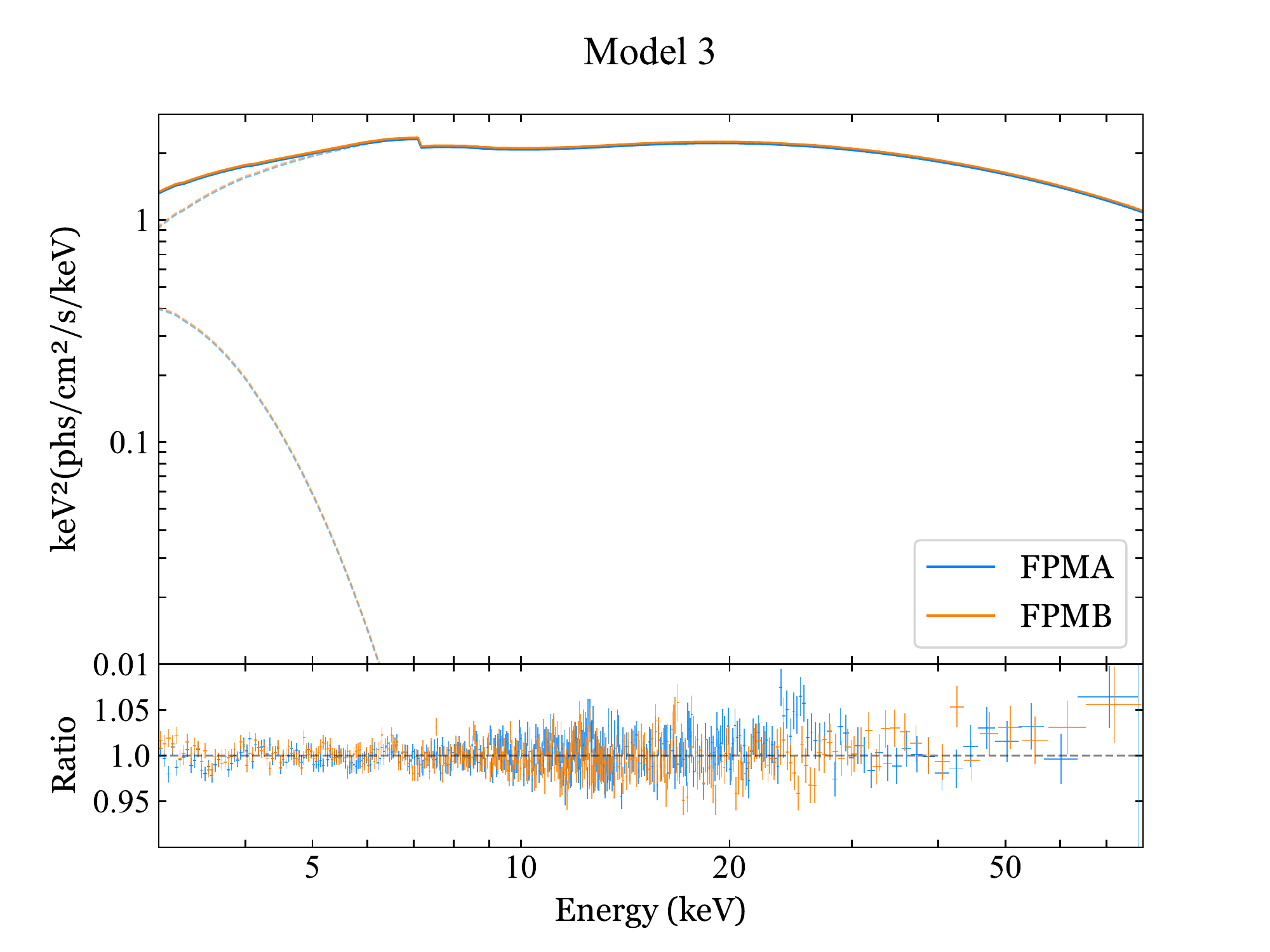}
		\includegraphics[width=0.48\textwidth,trim=0.8cm 0.0cm 1.0cm 0.0cm,clip]{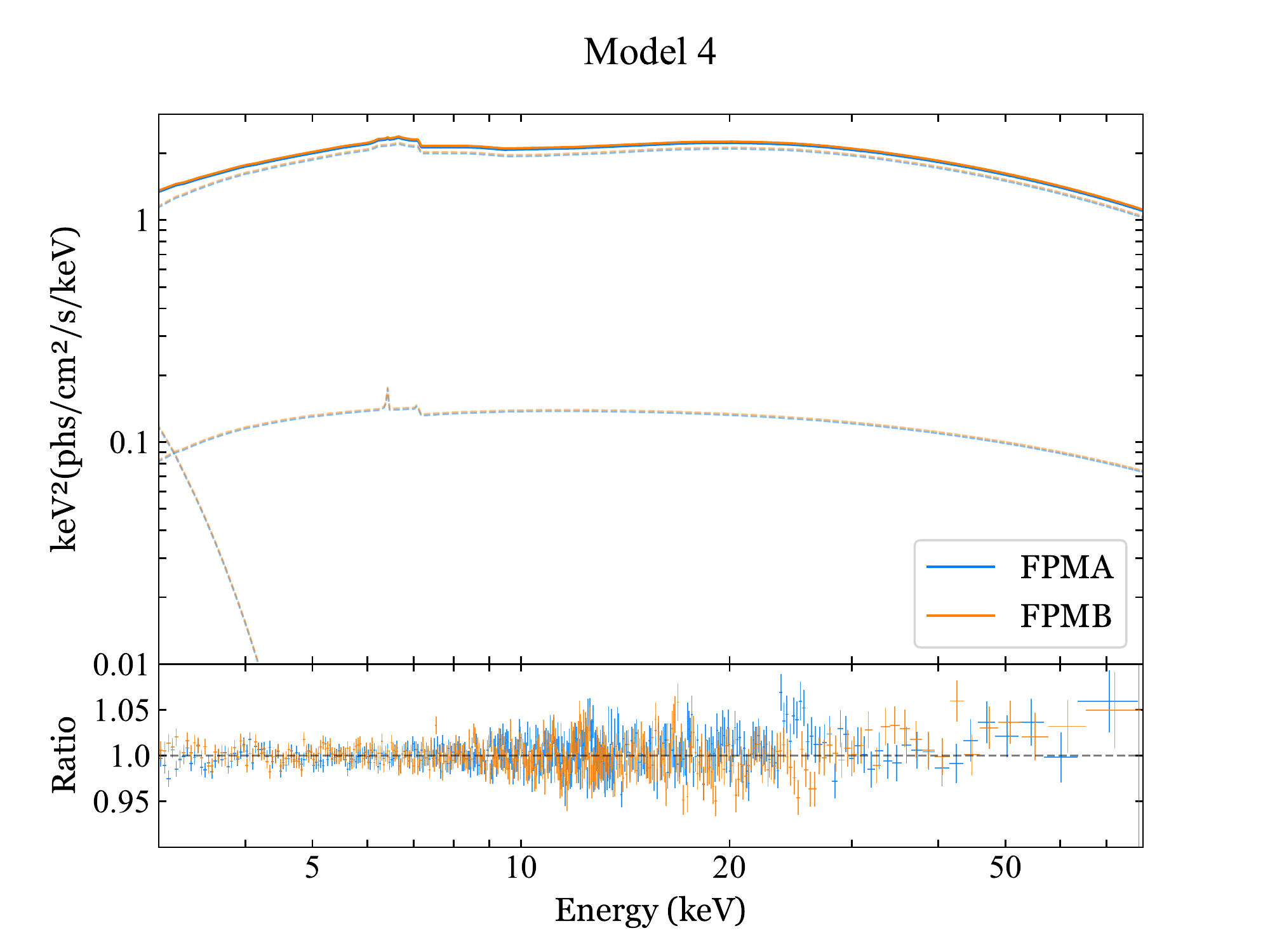} \\
	\end{center}
	\vspace{-0.4cm}
	\caption{Best-fit model and data to best-fit model ratio for Models~1-4. 
	%The upper-left panel represents Model~1; the upper-right panerl represents Model~2; the lower-left panel represents Model~3; the lower-right panel represents Model~4. 
	\label{bfp}}
\end{figure*}

In our first fit (Model~1), we fit the data assuming that the emissivity profile of the disk is described by a broken power law (so $p_2 = p_3$ and $R_\mathrm{br1} = R_\mathrm{br2}$). The best-fit values are reported in Tab.~\ref{bft}. The best-fit model and the data to best-fit model ratio are shown in the top-left panel in Fig.~\ref{bfp}. As we can see from Tab.~\ref{bft}, we find a very high emissivity index for the inner region of the accretion disk and an almost vanishing emissivity index for the outer part. Such an emissivity profile may be generated by a corona covering a large portion of the accretion disk~\cite{Miniutti:2003yd,Wilkins:2015nfa,Riaz:2020svt,Liu:2021tyw} and the data may prefer a twice broken power law. As Model~2, we thus fit the spectrum  with a twice broken power law. The best-fit values are reported in the third column of Tab.~\ref{bft} and the best-fit model and the data to best-fit model ratio are shown in the top-right panel in Fig.~\ref{bfp}. We do not see any improvement in the fit and $R_\mathrm{br2}$ is stuck at the outer edge of the accretion disk. Unfortunately, for the outer edge of the disk we have already chosen the maximum value allowed by the model and we cannot fit the data with a larger disk; such a value for $R_\mathrm{br2}$ should thus be understood as a lower limit. As Model~3, we reconsider an emissivity profile described by a broken power law, but this time we freeze the emissivity index of the outer region of the disk to 3, which is the value normally expected for the outer emissivity index when the corona is compact. The results are still shown in Tab.~\ref{bft} and Fig.~\ref{bfp}, but the fit is clearly worse with $\Delta\chi^2 = + 65$ with respect to Model~1. Last, we consider the possibility of the presence of a distant cold reflector and we add {\tt xillver}~\cite{Garcia:2013oma} to the total model. As shown in Tab.~\ref{bft}, these data clearly do not require any distant reflector.

To compare the quality of the fits of Models~1-4, we consider the Akaike information criterion (AIC) \cite{akaike}, which is a more robust method than the comparison of the minima of $\chi^2$. Since the sample size is not large with respect to the number of free parameters, we calculate the Akaike information criterion corrected for small sample sizes (AICc)~\cite{bookaicc}
\be
{\rm AICc} = \chi^2_{\rm min} + 2 N_p + \frac{2 N_p \left( N_p + 1 \right)}{\left( N_b - N_p - 1 \right)} \, ,
\ee
where $N_p$ is the number of free parameters and $N_b$ is the number of bins. The AICc values for Models~1-4 are reported in the last row of Tab.~\ref{bft}. The best model is that with the lowest AICc (Model~1). As a general and empirical rule, models with $\Delta{\rm AICc}>5$ are less favored by the data and models with $\Delta{\rm AICc}>10$ can be ruled out and omitted from further consideration~\cite{bookaicc}. With such a criterion, we can rule out Model~3, while Model~2 and Model~4 are still acceptable and can be used to estimate our modeling uncertainties in the measurements of the parameters of the system.

\begin{table*}[]
         \renewcommand\arraystretch{1.5}
	\centering
	\vspace{0.5cm}
	\begin{tabular}{lcccc}
		\hline\hline
		Model            & 1 & 2 & 3 & 4 \\
		\hline
		{\tt tbabs}   \\
		$ N_{\rm H}/10^{22}$~cm$^{-2}$ &  $4.3_{-0.4}^{+0.3}$   & $4.2_{-0.4}^{+0.5}$ & $5.8_{-0.4}^{+0.4}$&$4.2_{-0.4}^{+0.3}$  \\
		\hline
		{\tt diskbb}  \\ 
		$ T_{\rm in} $ [keV] &   $0.31_{-0.08}^{+0.10}$  &$0.31_{-0.09}^{+0.09}$  &$0.497_{-0.012}^{+0.013}$  &$0.31_{-0.09}^{+0.09}$   \\ 
		Norm$/10^{5}$ &  $1.4_{-1.2}^{+10}$  & $1.1_{-0.9}^{+8}$ & $0.084_{-0.009}^{+0.018}$ & $1.2_{-1.1}^{+8}$    \\
		\hline
		{\tt relxillion\_nk}  \\ 
		$ p_{1} $&   $10.0_{-2.4}$  & $10.0_{-2.3}$ & $10.0_{-0.4}$ & $10.0_{-2.3}$   \\ 
		$ p_{2} $&  $0.5^{+0.7}$  & $0.4^{+0.6}$  & $3^*$ &  $0.5^{+0.7}$ \\ 
		$ p_{3} $&  -- & $3^*$ & -- & -- \\ 
		$ R_{\rm br1} $ [$r_{\rm g}$]&    $5.2_{-1.5}^{+2.1}$  & 	$5.5_{-1.5}^{+2.9}$ & $2.86_{-0.10}^{+0.16}$ & $5.1_{-2.3}^{+2.8}$ \\ 
		$ R_{\rm br2} $ [$r_{\rm g}$]&    -- & $900_{-490}$ & -- & -- \\ 
		$ a_* $&  $0.998_{-0.004}$  & $0.998_{-0.004}$ & $0.998_{-0.007}$ &  $0.998_{-0.004}$\\ 
		$ i $ [deg]&  $78.2_{-1.2}^{+0.6}$  & $78.0_{-1.2}^{+0.9}$ & 	$69.3_{-0.9}^{+1.6}$ & $78.2_{-1.1}^{+0.7}$ \\ 
		$\Gamma$ & $2.04_{-0.09}^{+0.05}$ & $2.01_{-0.06}^{+0.05}$ & $1.84_{-0.03}^{+0.02}$ & $2.02_{-0.07}^{+0.07}$ \\
		$\log\xi_{\rm in}$ [$\rm erg \cdot \rm cm \cdot \rm s^{-1}$]&   $3.00_{-0.14}^{+0.08}$ & $3.05_{-0.19}^{+0.07}$ & $3.61_{-0.07}^{+0.11}$ & $3.03_{-0.17}^{+0.09}$ \\ 
		$ A_{\rm Fe} $&   $1.5_{-0.5}^{+0.4}$ & $1.5_{-0.5}^{+0.5}$ & $2.7_{-0.3}^{+0.7}$ & $1.5_{-0.5}^{+0.5}$  \\ 
		$E_{\rm cut}$ [keV]  & $110_{-25}^{+18}$ & $103_{-6}^{+22}$ & $80_{-7}^{+5}$ & $106_{-12}^{+19}$ \\
		$R_{\rm f}$ & $0.237_{-0.015}^{+0.019}$ & $0.222_{-0.014}^{+0.014}$  & $0.25_{-0.04}^{+0.06}$  & $0.25_{-0.02}^{+0.12}$ \\
		$\alpha _ \xi$ & $0.19_{-0.05}^{+0.06}$ & $0.19_{-0.03}^{+0.04}$ & $0.00^{+0.07}$ & $0.19_{-0.04}^{+0.06}$\\
		$q$ &   $0.57_{-0.7}^{+0.11}$ & $0.57_{-0.5}^{+0.10}$ & $1.89_{-0.21}$ & $0.57_{-0.5}^{+0.10}$  \\ 
		Norm$/10^{-2}$ &  $2.58_{-0.34}^{+0.18}$ & $2.48_{-0.21}^{+0.19}$ & $1.72_{-0.08}^{+0.16}$ & $2.3_{-0.5}^{+0.4}$  \\
		\hline
		{\tt xillver} \\ 
		Norm$/10^{-3}$ & --  & -- & --  & $2^{+5}$   \\ 
		\hline
		{\tt constant} \\ 
		FPMA & $1^*$ & $1^*$ & $1^*$ & $1^*$ \\
		FPMB & $1.0152_{-0.0014}^{+0.0014}$ & $1.0152_{-0.0014}^{+0.0014}$ & $1.0152_{-0.0014}^{+0.0014}$ & $1.0152_{-0.0014}^{+0.0014}$ \\
		\hline
		$\chi^2/\nu $ & $\quad 2659.62/2599 \quad$ & $\quad 2659.59/2598 \quad$ & $\quad 2724.98/2600 \quad$ & $\quad 2659.41/2598 \quad$ \\ 
		& $=1.02332$ & $=1.02371$ & $=1.04807$ & $=1.02364$\\
		\hline
		AICc & {2693.86} & {2695.85} &  {2757.19} & {2695.67} \\
		\hline\hline
	\end{tabular}
	\vspace{0.2cm}
	\caption{Best-fit table of Models~1-4. The reported uncertainties correspond to the 90\% confidence level for one relevant parameter ($\Delta\chi^2=2.71$). $^*$ means the value of the parameter is frozen during the fit. When there is no lower/upper uncertainty, the boundary of the range in which the parameter is allowed to vary is within the 90\% confidence limit. \label{bft}}
\end{table*}

\begin{figure*}[t]
	\begin{center}
		\includegraphics[width=0.8\textwidth,trim=0cm 0cm 0cm 0cm,clip]{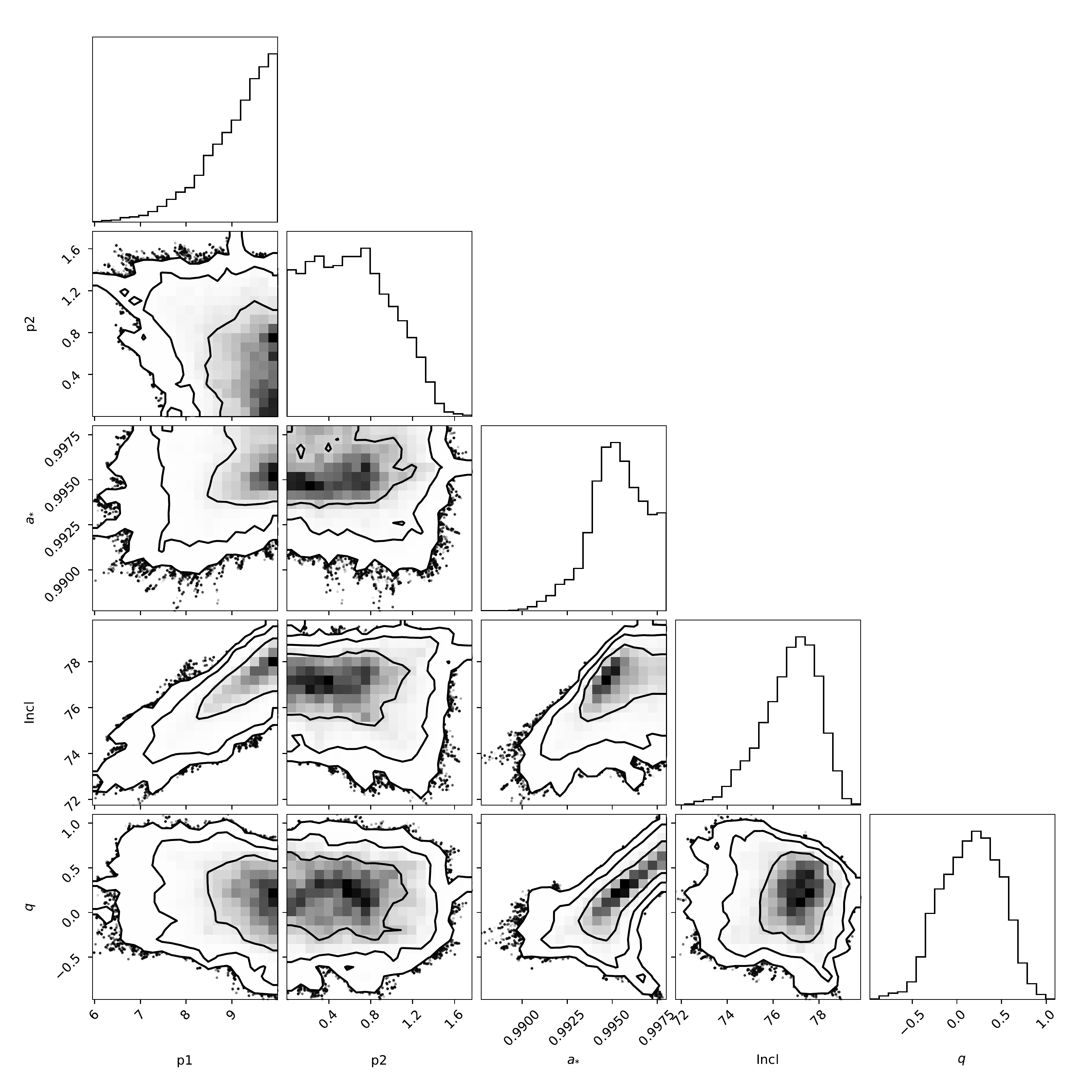}
	\end{center}
	\vspace{-0.8cm}
	\caption{Corner plot for $p_1$, $p_2$, $a_*$, $i$, and $q$ in Model~1 after the MCMC run. The 2D plots show the 1-, 2-, and 3-$\sigma$ confidence level contours.
	\label{f-mcmc}}
\end{figure*}

Since we simultaneously fit several parameters, it is useful to check possible parameter degeneracy. To do this, we run a Markov Chain Monte-Carlo (MCMC) analysis of Model~1. The corner plot showing possible correlations among the parameters $p_1$, $p_2$, $a_*$, $i$, and $q$ is reported in Fig.~\ref{f-mcmc}. The deformation parameter $q$ is clearly correlated to the spin parameter $a_*$, as both regulate the location of the ISCO radius. However, thanks to the high quality data of this spectrum, we can simultaneously constrain $q$ and $a_*$. On the other hand, we do not see any apparent correlation of the measurement of $q$ with the estimates of $p_1$, $p_2$, and $i$.

%%%%%%%%%%%%%%%%%%%%%%%%%%%%%%%

\section{Discussion and conclusions \label{S5}}

As discussed in Section~\ref{S3}, it is extremely important to select the right sources and observations to get precise and accurate measurements of the spacetime metric. From previous studies, we know that the 2019 \textsl{NuSTAR} spectrum of EXO~1846--031 is a good choice for our goal. The observation has been already analyzed and extensively discussed in the literature~\cite{exoN,Abdikamalov:2021rty,Abdikamalov:2021ues,Yu:2021xen}: the systematic effects seem to be well understood and the final measurement of the deformation parameter is robust. With this spirit, here we have reported the analysis of this single but well-understood spectrum. There are a few other well understood spectra of other sources, and their analysis can presumably lead to similar constraints on $q$~\cite{Bambi:2022dtw}. The situation is different from, for example, black hole spin measurements, where one wants to understand the spin distribution of the whole black hole population to figure out the gravitational collapse process behind the formation of black holes while the measurement of a single source cannot address general questions.

Model~1 is the simplest model and fits the data well. Our constraint on $q$ is (90\% confidence level, only statistical uncertainty)
\be\label{eq-constraint}
-0.1 < q < 0.7 \, ,
\ee
and therefore our analysis is consistent with the hypothesis that the gravitationally collapsed object in the X-ray binary EXO~1846--031 is a Kerr black hole (for which $q=0$). However, our analysis does not exclude $q = \mathcal{O}(0.1)$  and therefore natural values\footnote{Here we adopt the point of view widely accepted in theoretical physics, where the dimensionless fundamental parameters of a theory should be expected all of order 1. Parameters with values much larger or smaller than 1 are ``unnatural'' (and presumably they are not fundamental, but derived quantities from a more fundamental theory).} of the deformation parameter are allowed. We note that positive values of $q$ implies that a non-rotating source is oblate, which is more physically realistic than a prolate (i.e. $q<0$) one. In the case of a rotating source, $q > 0$ implies that the object is more oblate than a Kerr black hole and $q<0$ implies that it is more prolate than a Kerr black hole (but not necessarily that it is a prolate object).

The constraint in Eq.~(\ref{eq-constraint}) can be rewritten in terms of a constraint on the mass quadrupole of the source. For $a_* = 0.998$, we find
\be\label{eq-constraint2}
-0.955 < M_2/M^3 < -1.131 \, ,
\ee
while for a Kerr black hole we have $M_2/M^3 = -0.996$. Objects with equations of state similar to neutron stars would be significantly more oblate and the value of $M_2/M^3$  would be in the range $-3$ to $-10$~\cite{Laarakkers:1997hb}.

From Model~1, we find that the emissivity profile is very steep around the central object and almost flat at larger radii. While this is not the emissivity profile expected from a compact corona, it is common in Galactic black holes; see, for instance, the discussion in \cite{Liu:2021tyw} and references therein. If we try to fit the data with a twice broken power law (Model~2) or by adding a non-relativistic reflection component (Model~4), we do not see any significant difference: the value of the second breaking radius would be large and the normalization of the non-relativistic reflection component would be very low. The estimate of the model parameters are thus consistent with the measurements inferred with Model~1. If we model the emissivity profile with a broken power law and we impose that the outer emissivity index is 3, the estimate of some model parameters would be different (and we find that the spacetime significantly deviates from the Kerr solution!), but the fit is definitively worse and Model~3 can be ruled out ($\Delta{\rm AICc} = 63$ with respect to Model~1).

Systematic uncertainties related to simplifications in our theoretical model to fit the data are under control and cannot significantly change our results in Eq.~(\ref{eq-constraint}) and Eq.~(\ref{eq-constraint2}). In Appendix~\ref{app-nustar}, we show the results of some simulations created with models with a disk of finite thickness, a non-vanishing electron density gradient, and a corona of specific geometry. We fit the simulated data with our theoretical model and we do not see any significant bias in the estimate of most parameters, in particular in the estimate of the black hole spin and the parameter $q$.

We are not aware of other tests of the $\delta$-Kerr metric published in the literature and observational constraints on the deformation parameter $q$, though models for the shadow and quasinormal modes have been studied in Refs. \cite{Li2022,metric}. 
Since the $\delta$-Kerr spacetime is an exact vacuum solutions of the field equations in general relativity which relates to the Kerr black hole through the variation of one continuous parameter with a clear physical interpretation, we would argue that experimental tests to constrain the allowed values of $q$ from observations are important towards a possible resolution of the Kerr hypothesis.
We could certainly constrain $q$ from the available gravitational wave data from the LIGO-Virgo-KAGRA Collaboration following the approach employed in Refs.~\cite{Cardenas-Avendano:2019zxd,Shashank:2021giy} for testing other non-Kerr metrics. The deformation parameter $q$ may also be constrained from the available mm black hole images from the Event Horizon Telescope Collaboration (see, e.g., Ref.~\cite{Vagnozzi:2022moj}). While these three techniques (X-ray, gravitational waves, and black hole imaging) are sensitive to different relativistic effects, in general, X-ray tests are those that can provide the most stringent constraints on possible deviations from the Kerr solution, while gravitational wave constraints are normally a bit weaker and black hole imaging constraints are more than an order of magnitude weaker; see, for example, Ref.~\cite{Bambi:2022dtw}. This is the typical situation with the current data. However, gravitational wave constraints are expected to improve quickly in the coming years.

Concerning X-ray tests, the constraint reported in the present work is likely close to the best that we can do today. Somewhat more stringent constraints may be obtained from sources in which we can test the Kerr metric from the simultaneous analysis of the reflection features and the thermal spectrum, as done in Refs.~\cite{Tripathi:2020dni,Zhang:2021ymo,Tripathi:2021rqs}. This is not possible for the \textsl{NuSTAR} spectrum analyzed here because the thermal component is too weak and we do not have independent measurements of the mass and distance of the compact object. More stringent constraints on the deformation parameter $q$ require higher quality data, which will be available from the next generation of X-ray missions, starting from \textsl{eXTP}~\cite{eXTP:2016rzs}, which is currently scheduled to be launched in 2027. The analysis of a simulated observation of a source like EXO~1846--031 with the LAD instrument on board \textsl{eXTP} is reported in Appendix~\ref{app-extp} and we find that we can improve the constraint on $q$ in Eq.~(\ref{eq-constraint}) by an order of magnitude.

%%%%%%%%%%%%%%%%%%%%%%%%%%%%%%%

\vspace{0.5cm}

{\bf Acknowledgments --}
This work was supported by the National Natural Science Foundation of China (NSFC), Grant No.~12250610185, 11973019, and 12261131497, the Natural Science Foundation of Shanghai, Grant No. 22ZR1403400, the Shanghai Municipal Education Commission, Grant No. 2019-01-07-00-07-E00035, and Fudan University, Grant No. JIH1512604. S.R. acknowledges the support from the China Postdoctoral Science Foundation, Grant No. 2022M720035. D.M. acknowledges support from Nazarbayev University Faculty Development Competitive Research Grant No. 11022021FD2926.

%%%%%%%%%%%%%%%%%%%%%%%%%%%%%%%

\appendix

\section{Simulated observations with  \textsl{NuSTAR} }\label{app-nustar}

In this appendix, we want to show that simplifications in our theoretical model to fit the data should not have a significant impact on the estimate of the parameters of the object, so our results in Eq.~(\ref{eq-constraint}) and Eq.~(\ref{eq-constraint2}) are robust. We simulate three observations of a source like EXO~1846--031 with \textsl{NuSTAR}, assuming $q=0$ (Kerr metric) and the best-fit values of Model~1 in Tab.~\ref{bft} as input parameters. The exposure time of each simulation is 30~ks. We have to run simulations and we cannot use these models to fit the \textsl{NuSTAR} spectrum of EXO~1846--031 because we do not have these models for the $\delta$-Kerr metric and their construction would be beyond the scope of the present work.

In our analysis of EXO~1846--031, the electron density is supposed to be constant over the disk. In our first simulation, we replace {\tt relxillion\_nk} with {\tt relxilldgrad\_nk}, where the electron density profile is modeled with a power law and the ionization profile is calculated self-consistently from the emissivity and the electron density at every radius~\cite{Abdikamalov:2021ues}. We assume that the electron density at the inner edge of the accretion disk is $n_{\rm in} = 10^{17}$~cm$^{-3}$ and the electron density profile scales as $1/r^2$.

In our second simulation, we consider a disk of finite thickness, while our analysis of EXO~1846--031 is based on a model with an infinitesimally thin disk. We employ the flavor of {\tt relxill\_nk} described in Ref.~\cite{Abdikamalov:2020oci} and we assume that the Eddington-scaled mass accretion rate is $\dot{m} = 0.2$.

Last, in our third simulation we employ a corona of specific geometry. We use the model described in Ref.~\cite{Riaz:2020svt}, where the corona is a disk at a height $h_{\rm corona}$ above the accretion disk and with a radius $R_{\rm corona}$. We assume $h_{\rm corona} = 5~r_{\rm g}$ and $R_{\rm corona}= 15~r_{\rm g}$, where $r_{\rm g}$ is the gravitational radius of the source.

\begin{figure*}[t]
	\begin{center}
		\includegraphics[width=1.0\textwidth,trim=2.0cm 0.0cm 3.0cm 0.0cm,clip]{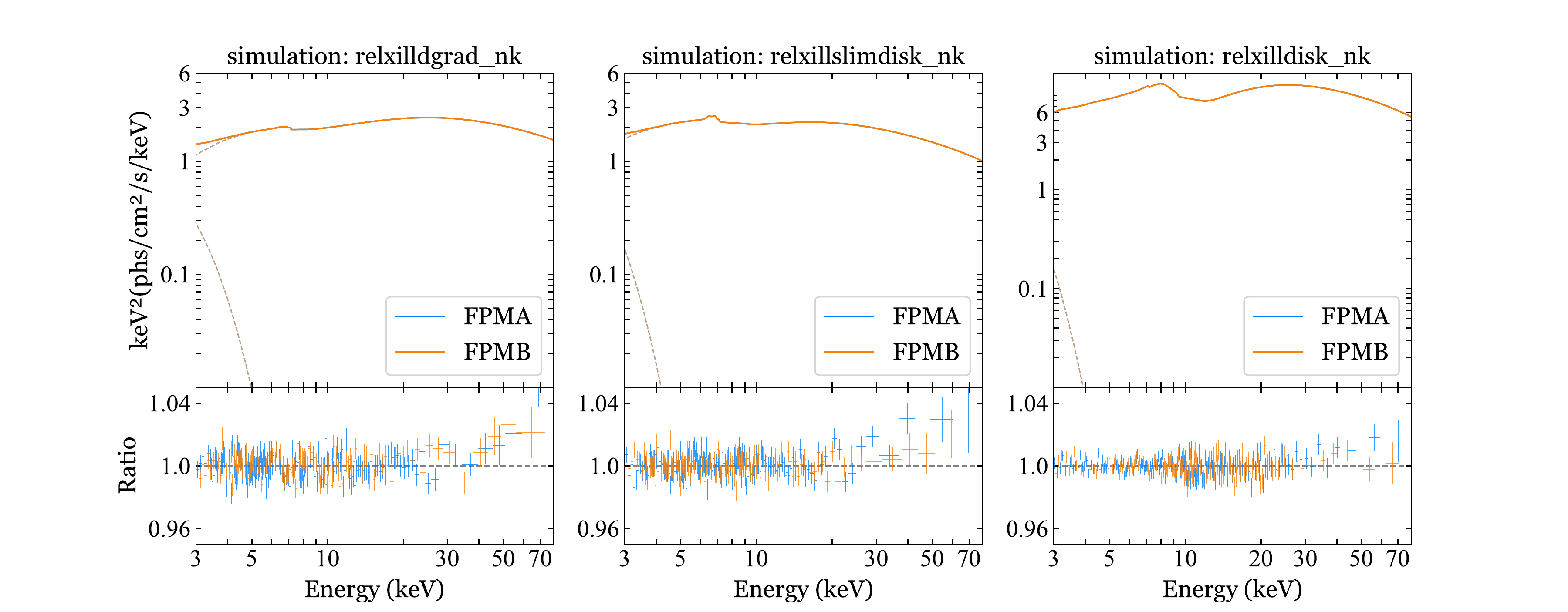}
	\end{center}
	\vspace{-0.4cm}
	\caption{Best-fit model and data to best-fit model ratio for the simulated observation with \textsl{NuSTAR}.
	\label{bfpa}}
\end{figure*}

We fit the three simulations with Model~1 with the parameter $q$ free and the emissivity profile modeled by a broken power law. The results of our fits are reported in Tab.~\ref{bfta} and Fig.~\ref{bfpa}. We see that the iron abundance is significantly overestimated in the first simulation in which the input model has a non-vanishing electron density gradient. However, in general, we can recover the correct input parameters; that is, the simplifications in our {\tt relxillion\_nk} used to analyze EXO~1846--031 should not affect the conclusions of our work. In particular, we do not see any significant impact on the estimates of the black hole spin parameter $a_*$ and on the deformation parameter $q$.

\begin{table*}[t]
         \renewcommand\arraystretch{1.5}
	\centering
	\vspace{0.5cm}
	\begin{tabular}{lcccccc}
		\hline\hline
		            & \hspace{0.5cm} Input \hspace{0.5cm} & \hspace{0.5cm} Fit \hspace{0.5cm}& \hspace{0.5cm} Input \hspace{0.5cm} & \hspace{0.5cm} Fit \hspace{0.5cm} &\hspace{0.5cm} Input \hspace{0.5cm} & \hspace{0.5cm} Fit \hspace{0.5cm}  \\
		\hline
		{\tt tbabs}   \\
		$ N_{\rm H}/10^{22}$~cm$^{-2}$ &  $4.3$   & $4.3^*$ & $4.3$ & $4.3^*$ & $4.3$ & $4.3^*$  \\
		\hline
		{\tt diskbb}  \\ 
		$ T_{\rm in} $ [keV] &   $0.31$  & $0.400_{-0.020}^{+0.020}$ & $0.31$ & $0.290_{-0.070}^{+0.020}$ & $0.31$ & $0.25_{-0.04}^{+0.06}$ \\ 
		%Norm$/10^{5}$ &  $1.4$  & $1.1_{-0.9}^{+8}$  \\
		\hline
		&{\tt relxilldgrad\_nk} & & {\tt relxillslimdisk\_nk} && {\tt relxilldisk\_nk}  \\ 
		$h_{\rm corona}~[r_{\rm g}]$ & $-$  & $-$ & $-$ & $-$ & $5$ & $-$  \\
		$R_{\rm corona}~[r_{\rm g}] $ & $-$ & $-$ & $-$ & $-$ & $15$ & $-$\\
		$ p_{1} $&   $10$  & $10_{-0.4}$ & $10$ & $10_{-2.0}^{}$ & $-$ &  $7.0_{+2.1}^{-7.0}$ \\ 
		$ p_{2} $&  $0.5$  & $0.030_{-0.020}^{+1.000}$ & $0.5$ & $0.0004_{-0.0002}^{+0.3713}$ & $-$ & $3.325_{-0.036}^{+0.035}$ \\ 
		$ R_{\rm br1} $ [$r_{\rm g}$]&    $5.2$  & 	$9.0_{-0.7}^{+7.0}$ & $5.2$  & $7.2_{-2.5}^{+2.7}$ & $-$ & $12.0_{-1.1}^{+0.5}$\\ 
		$ a_* $&  $0.998$  & $0.998_{-0.003}$ & $0.998$ & $0.993_{-0.002}^{}$ & $0.998$ & $0.998_{-0.010}^{}$ \\ 
		$ i $ [deg]&  $78.2$  & $69.4_{-1.5}^{+0.5}$ & $78.2$ & $69.7_{-1.3}^{+1.1}$ & $78.2$ & $78.43_{-0.16}^{+0.16}$ \\ 
		$\Gamma$ & $2.04$ & $1.820_{-0.004}^{+0.025}$ & $2.04$ & $2.034_{-0.040}^{+0.050}$ & $2.04$ & $2.041_{-0.007}^{+0.007}$  \\
		$\log\xi_{\rm in}$ [$\rm erg \cdot \rm cm \cdot \rm s^{-1}$]&   $3$ & $2.800_{-0.020}^{+0.050}$ & $3$ & $3.017_{-0.012}^{+0.035}$ & $3$ & $3.0024_{-0.0013}^{+0.0015}$ \\ 
		$ A_{\rm Fe} $&   $1.5$ & $5.00_{-0.20}^{+0.40}$ & $1.5$ & $1.50_{-0.30}^{+0.23}$ & $1.5$ & $1.442_{-0.013}^{+0.011}$ \\ 
		$E_{\rm cut}$ [keV]  & $110$ & $104.0_{-2.0}^{+2.0}$ & $110$ & $104_{-5}^{+8}$ & $110$ & $140.7_{-1.1}^{+1.4}$  \\
		$R_{\rm f}$ & $0.237$ & $0.248_{-0.010}^{+0.006}$ & $0.237$ & $0.238_{-0.036}^{+0.045}$ & $0.237$ & $3.68_{-0.20}^{+0.20}$ \\
		${\rm log}~n_{\rm in}~[\rm cm^{-3}]$ & $17$ & $15^*$ & $15$ & $15^*$ & $15$ & $15^*$ \\
		$\alpha_{n}$ & $2$ & $0^*$ & $-$ & $-$ & $-$ & $-$\\
		$\dot{m}$ & $-$ & $-$ & $0.2$ & $0^*$ & $-$ & $-$ \\
		$q$ & $-$ & $0.40_{-0.50}^{+0.04}$ & $-$ & $0.013_{-0.845}^{+0.230}$ & $-$ & $0.32_{-0.32}^{+0.76}$ \\
		\hline
		{\tt constant} \\ 
		FPMA & $-$ & $1^*$ & $-$ & $1^*$ & $-$ & $1^*$ \\
		FPMB & $-$ & $0.9997_{-0.0012}^{+0.0012}$ & $-$ & $1.0007_{-0.0011}^{+0.0011}$ & $-$ & $0.9999_{-0.0001}^{+0.0001}$ \\
		%Norm$/10^{-2}$ &  $5.58$ & $2.48_{-0.21}^{+0.19}$ & $1.72_{-0.08}^{+0.16}$ & $2.3_{-0.5}^{+0.4}$  \\
		\hline
		$\chi^2/\nu $ &  & $2958.38/2906$ & & $2673.01/2713$ & & $3599.87/3446$ \\ 
		& & $=1.01802$ && $=0.98525$ & & $=1.04465$   \\
		\hline\hline
	\end{tabular}
	\vspace{0.2cm}
	\caption{Best-fit table of simulated 30~ks observation of EXO~1846--031 with \textsl{NuSTAR}. The reported uncertainties correspond to the 90\% confidence level for one relevant parameter ($\Delta\chi^2=2.71$). * means the value of the parameter is frozen during the fit. } \label{bfta}
\end{table*}

\section{Simulated observation with LAD/\textsl{eXTP}}\label{app-extp}

\begin{figure}[t]
	\begin{center}
		\includegraphics[width=0.48\textwidth,trim=0.1cm 0.0cm 1.0cm 0.0cm,clip]{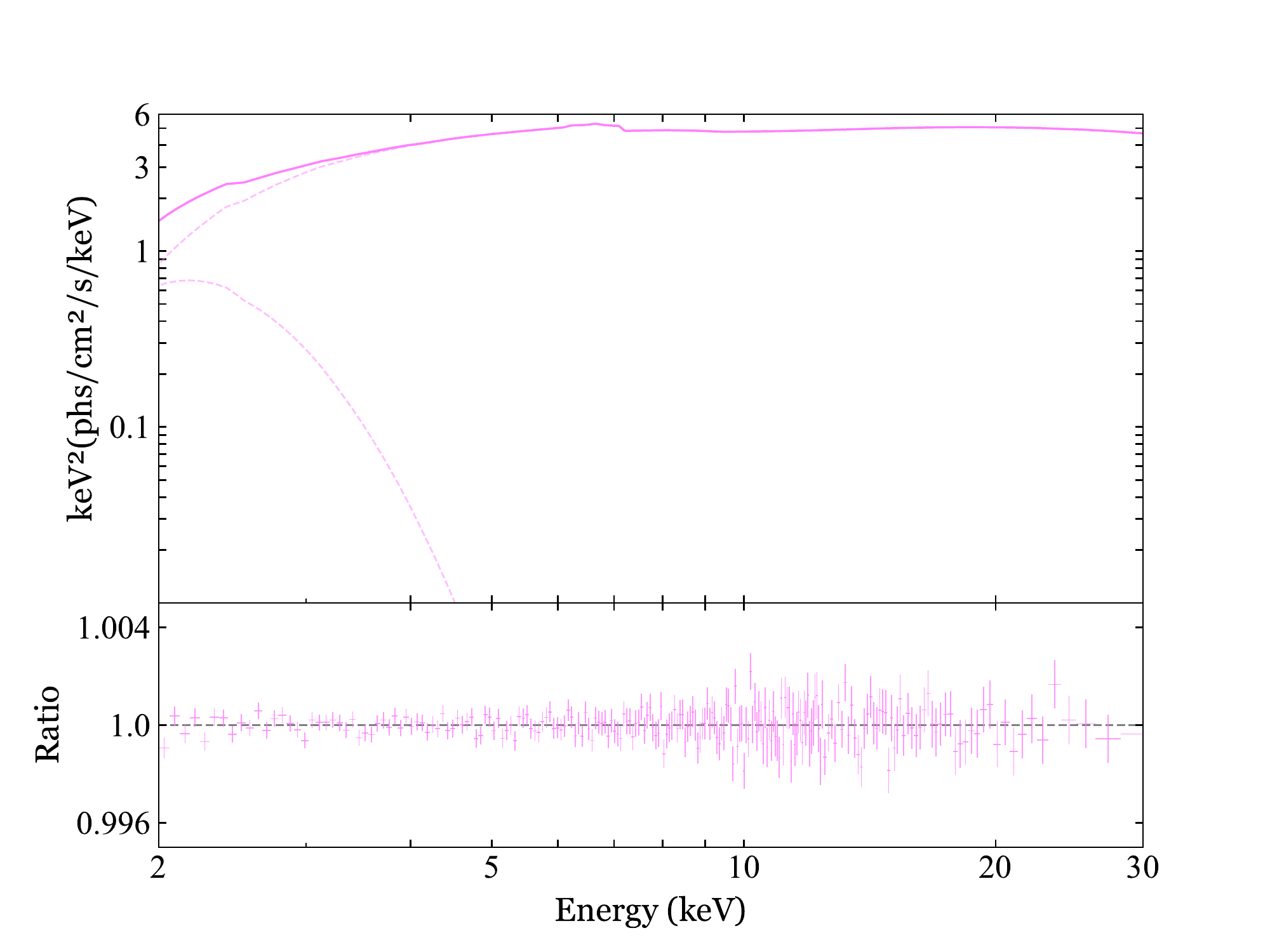}
	\end{center}
	\vspace{-0.4cm}
	\caption{Best-fit model and data to best-fit model ratio for the simulated observation with LAD/\textsl{eXTP}.
	\label{bfp2}}
\end{figure}

As discussed in Section~\ref{S5}, the constraint on $q$ inferred in the present work is likely close to the best that we can do today. More stringent constraints will be possible with the next generation of X-ray observatories. To be more quantitative, we simulate an observation of a source like EXO~1846--031 with the LAD instrument on board \textsl{eXTP}. We assume a 30~ks observation and we calculate a spectrum with Model~1, employing the best-fit values of Model~1 in Tab.~\ref{bft} as input parameters but assuming that the source is a Kerr black hole and thus $q=0$. From the analysis of this simulated observation, we find the measurements reported in Tab.~\ref{bft2}. In particular, the measurement of the deformation parameter $q$ is (90\% confidence level)
\be
-0.05 < q < 0.02 \, ,
\ee
which is roughly an order of magnitude better than our measurement with \textsl{NuSTAR} in Eq.~(\ref{eq-constraint}). Fig.~\ref{bfp2} shows the best-fit model and the data to the best-fit model ratio.

\begin{table}[t]
         \renewcommand\arraystretch{1.5}
	\centering
	\vspace{0.5cm}
	\begin{tabular}{lcc}
		\hline\hline
		            & \hspace{0.5cm} Input \hspace{0.5cm} & \hspace{0.5cm} Fit \hspace{0.5cm} \\
		\hline
		{\tt tbabs}   \\
		$ N_{\rm H}/10^{22}$~cm$^{-2}$ &  $4.3$   & $4.267_{-0.002}^{+0.005}$  \\
		\hline
		{\tt diskbb}  \\ 
		$ T_{\rm in} $ [keV] &   $0.31$  & $0.3083_{-0.0003}^{+0.0005}$ \\ 
		%Norm$/10^{5}$ &  $1.4$  & $1.1_{-0.9}^{+8}$  \\
		\hline
		{\tt relxillion\_nk}  \\ 
		$ p_{1} $&   $10$  & $9.96_{-0.21}$ \\ 
		$ p_{2} $&  $0.5$  & $0.49_{-0.03}^{+0.05}$ \\ 
		$ R_{\rm br1} $ [$r_{\rm g}$]&    $5.2$  & 	$5.23_{-0.18}^{+0.17}$ \\ 
		$ a_* $&  $0.998$  & $0.998_{-0.004}$ \\ 
		$ i $ [deg]&  $78.2$  & $78.10_{-0.08}^{+0.07}$ \\ 
		$\Gamma$ & $2.04$ & $2.0369_{-0.0013}^{+0.0025}$ \\
		$\log\xi_{\rm in}$ [$\rm erg \cdot \rm cm \cdot \rm s^{-1}$]&   $3$ & $3.005_{-0.011}^{+0.007}$ \\ 
		$ A_{\rm Fe} $&   $1.5$ & $1.53_{-0.03}^{+0.05}$ \\ 
		$E_{\rm cut}$ [keV]  & $110$ & $109.2_{-0.9}^{+1.0}$ \\
		$R_{\rm f}$ & $0.237$ & $0.237_{-0.003}^{+0.002}$ \\
		$\alpha _ \xi$ & $0.19$ & $0.191_{-0.002}^{+0.003}$ \\
		$q$ &   $0$ & $-0.003_{-0.051}^{+0.012}$ \\ 
		%Norm$/10^{-2}$ &  $5.58$ & $2.48_{-0.21}^{+0.19}$ & $1.72_{-0.08}^{+0.16}$ & $2.3_{-0.5}^{+0.4}$  \\
		\hline
		$\chi^2/\nu $ &  & $1267.56/1294$ \\ 
		& & $=0.97957$  \\
		\hline\hline
	\end{tabular}
	\vspace{0.2cm}
	\caption{Best-fit table of a simulated 30~ks observation of EXO~1846--031 with \textsl{NuSTAR}. The reported uncertainties correspond to the 90\% confidence level for one relevant parameter ($\Delta\chi^2=2.71$). } \label{bft2}
\end{table}

%%%%%%%%%%%%%%%%%%%%%%%%%%%%%%%

\end{document}